\documentclass[aps,showpacs,preprintnumbers,amsmath,nofootinbib,amssymb]{revtex4}
\usepackage[dvips]{epsfig}

\newcommand{\nc}[1]{\newcommand{#1}}

\nc{\its}[1]{\itshape #1 \upshape}
\nc{\mc}[3]{\multicolumn{#1}{#2}{#3}}

\nc{\bc}{\begin{center}}
\nc{\ec}{\end{center}}
\nc{\ig}[1]{\bc \includegraphics{#1} \ec}

\nc{\bo}[1]{\mbox{\boldmath \( #1 \! \! \)  \unboldmath}}
\nc{\be}{\begin{eqnarray}}
\nc{\ee}{\end{eqnarray}}
\nc{\bew}{\begin{eqnarray*}}
\nc{\eew}{\end{eqnarray*}}
\nc{\bs}{\begin{subeqnarray}}   
\nc{\es}{\end{subeqnarray}}     
\nc{\nnn}{\nonumber \\}
\nc{\f}[2]{\frac{#1}{#2}}
\nc{\td}[2]{\f{d #1}{d #2}}
\nc{\pd}[2]{\f{\partial #1}{\partial #2}}
\nc{\suli}{\sum\limits}
\nc{\proli}{\prod\limits}
\nc{\ili}{\int\limits}
\nc{\sr}[2]{\stackrel{#1}{#2}}
\nc{\dps}{\displaystyle}
\nc{\ket}[1]{\left| #1 \right>}
\nc{\bra}[1]{\left< #1 \right|}
\nc{\bracket}[2]{\left< #1 \right| \left. \! #2 \right>}
\nc{\norm}[1]{\left\| #1 \right\|}
\nc{\lndm}[1]{\pd{^{#1} \ln{\det{M}}}{\mu^{#1}}}
\nc{\pdmm}[1]{M^{-1} \pd{^{#1} M}{\mu^{#1}}}
\nc{\pdm}{M^{-1}\pd{M}{\mu}}
\nc{\trac}[1]{\mbox{Tr}\left(#1\right)}
\nc{\hm}{\hat{m}}

\def\lsim{\raise0.3ex\hbox{$<$\kern-0.75em\raise-1.1ex\hbox{$\sim$}}}
\def\gsim{\raise0.3ex\hbox{$>$\kern-0.75em\raise-1.1ex\hbox{$\sim$}}}

\begin{document}

\title{The transition temperature in QCD}

\author{M. Cheng$^{\rm a}$, N. H. Christ$^{\rm a}$, S. Datta$^{\rm b}$, 
J. van der Heide$^{\rm c}$,
C. Jung$^{\rm b}$, F. Karsch$^{\rm b,c}$, O. Kaczmarek$^{\rm c}$,\\ 
E. Laermann$^{\rm c}$, R. D. Mawhinney$^{\rm a}$, C. Miao$^{\rm c}$,
P. Petreczky$^{\rm b,d}$, K. Petrov$^{\rm e}$,
C. Schmidt$^{\rm b}$ and T. Umeda$^{\rm b}$
}

\affiliation{
$^{\rm a}$ Physics Department,Columbia University, New York, NY 10027, USA\\
$^{\rm b}$Physics Department, Brookhaven National Laboratory, 
Upton, NY 11973, USA \\
$^{\rm c}$Fakult\"at f\"ur Physik, Universit\"at Bielefeld, D-33615 Bielefeld,
Germany\\
$^{\rm d}$ RIKEN-BNL Research Center, Brookhaven National Laboratory, 
Upton, NY 11973, USA \\
$^{\rm e}$Niels Bohr Institute, University of Copenhagen, 
Blegdamsvej 17, DK-2100 Copenhagen, Denmark\\
}

\date{\today}
\preprint{BNL-NT-06/27}
\preprint{BI-TP 2006/31}
\preprint{CU-TP-1158}

\begin{abstract}
We present a detailed calculation of the transition temperature in QCD with
two light and one heavier (strange) quark mass on lattices with temporal 
extent $N_\tau =4$ and $6$. Calculations with improved staggered fermions 
have been performed for various light to strange quark mass ratios in the
range, $0.05 \le \hm_l/\hm_s \le 0.5$, and with a strange quark mass fixed close 
to its physical value. From a combined extrapolation to the chiral 
($\hm_l\rightarrow 0$) 
and continuum ($aT \equiv 1/N_\tau \rightarrow 0$) limits we find 
for the transition temperature at the physical point $T_cr_0 = 0.457(7)$ 
where the scale is set by the Sommer-scale parameter $r_0$ defined as the
distance in the static quark potential at which the slope takes on the value,
$\left( {\rm d}V_{\bar{q}q}(r)/{\rm d}r\right)_{r=r_0} = 1.65/r_0^2$.
Using the currently best known value for $r_0$ this translates to a transition
temperature $T_c = 192(7)(4)$~MeV. 
The transition temperature in the chiral limit is about 3\% smaller. 
We discuss current ambiguities in the determination of $T_c$ in physical units
and also comment on the universal scaling behavior of thermodynamic quantities in the
chiral limit. 
\end{abstract}


\maketitle

\section{Introduction}
\label{intro}

It is by now well established that the properties of matter formed from 
strongly interacting elementary particles change drastically at high
temperatures. Quarks and gluons are no longer confined to move inside 
hadrons but organize in a new form of strongly interacting matter,
the so-called quark-gluon plasma (QGP). The transition from hadronic
matter to the QGP as well as properties of the high temperature phase
have been studied extensively in lattice calculations over recent 
years \cite{reviews}.
Nonetheless, detailed quantitative information on the transition and
the structure of the high temperature phase in the physical situation
of two light and a heavier strange quark (($2+1$)-flavor QCD) 
is rare \cite{peikert_pressure,Bernard04,Bernard,aoki}. 
In order to
relate experimental observables determined in relativistic heavy ion
collisions to lattice results, it is important to achieve 
good quantitative control, in calculations with physical quark masses,
over basic parameters that characterize the 
transition from the low to the high temperature phase of QCD. One of the
most fundamental quantities clearly is the transition temperature.
 
Many lattice calculations, performed in recent years, suggest that for
physical values of the quark masses, the transition to the high temperature 
phase of QCD is not a phase transition but rather a rapid crossover that occurs
in a small, well defined, temperature interval. In particular, the calculations 
performed with improved staggered fermion actions indicate a rapid
but smooth transition to the high temperature 
phase \cite{Bernard04,peikert};
distributions of the chiral condensate and Polyakov loop do not show
any evidence for metastabilities; and the volume dependence of observables 
characterizing the transition is generally found to be small. This allows 
one to perform studies of the transition in physical volumes of moderate size
which have already led to several calculations of the QCD transition temperature
for $2$ and $3$-flavor QCD on the lattice. 
A first chiral and continuum limit extrapolation of the transition temperature 
obtained in 
($2+1$)-flavor QCD with improved staggered fermions has been given recently
\cite{Bernard04}. A similar extrapolation of results obtained with
rather large quark masses has also been attempted for $2$-flavor QCD in
calculations performed with Wilson fermions \cite{Schierholz}.

In this paper we report on a new determination of the transition temperature
in QCD with almost physical light quark masses and a physical
value of the strange quark mass. Our calculations
have been performed with an improved staggered fermion action \cite{Heller} 
on lattices of temporal extent $N_\tau =4$ and $6$. We use the Rational
Hybrid Monte Carlo (RHMC) algorithm 
\cite{rhmc} to perform simulations with two light and a heavier strange quark. 
We will outline 
details of our calculational set-up in the next section. In section III we 
present our finite temperature calculations for the determination of the
transition point on finite lattices. Section IV is devoted to a discussion
of our zero temperature scale determination. We present our results on 
the transition temperature in Section V and conclude in Section VI.

\section{Lattice Formulation and calculational setup}
\label{setup}

We study the thermodynamics of QCD with two light quarks 
($\hm_l\equiv \hm_u= \hm_d$) and a heavier strange quark ($\hm_s$)
described by the QCD partition function which is discretized on a four
dimensional lattice of size $N_\sigma^3\times N_\tau$,
\begin{equation}
Z(\beta,\hat{m}_l,\hat{m}_s,N_\sigma,N_\tau) = 
\int \prod_{x,\mu} {\rm d}U_{x,\mu} 
\left( {\rm det}\; D(\hat{m}_l)\right)^{1/2}
\left( {\rm det}\; D(\hat{m}_s)\right)^{1/4} {\rm e}^{-\beta S_G (U)} \;\; . 
\label{partition}
\end{equation}  
Here we will use staggered fermions to discretize
the fermionic sector of QCD. 
The fermions have already been integrated out, which gives
rise to the determinants of the staggered fermion matrices, 
$D(\hm_l)$ and $D(\hm_s)$  for the contributions of two light and one 
heavy quark degree of freedom, respectively.
Moreover, $\beta = 6/g^2$ is the gauge coupling constant, $\hat{m}_{s,l}$
denote the dimensionless, bare quark masses in units of the lattice
spacing $a$, and $S_G$ is the
gauge action which is expressed
in terms of gauge field matrices $U_{x,\mu}\in SU(3)$ located on the links
$(x,\mu)\equiv (x_0,\bf{x},\mu)$ of the four dimensional lattice; 
$\mu=0,...,3$. 

In our calculations we use a tree level, ${\cal O}(a^2)$ improved gauge action,
$S_G$, which includes the standard Wilson plaquette term and the
planar 6-link Wilson loop. In the fermion sector, we use an improved staggered 
fermion action with 1-link and bended 3-link terms. The coefficient of the 
bended 3-link
term has been fixed by demanding a rotationally invariant quark propagator
up to ${\cal O}(p^4)$, which improves the quark dispersion relation at
${\cal O}(a^2)$. This eliminates ${\cal O}(a^2)$ corrections to the pressure
at tree level and leads to a strong reduction of cut-off effects
in other bulk thermodynamic observables in the infinite temperature limit,
as well as in ${\cal O}(g^2)$ perturbation theory \cite{Heller}. 
The 1-link term in the fermion action has
been `smeared' by adding a 3-link staple. This improves the flavor symmetry
of the staggered fermion action \cite{cheng}. We call this action the p4fat3
action. It has been used previously in studies of QCD thermodynamics on
lattices of temporal extent $N_\tau = 4$ with larger quark masses 
\cite{peikert_pressure,peikert}. We improve here on the old calculations 
performed with the p4fat3 action in several respects: 
(i) We perform calculations
with significantly smaller quark masses, which strongly reduces extrapolation
errors to the physical quark mass values; (ii) we obtain results for a 
smaller lattice cut-off by performing calculations on lattices with temporal 
extent $N_\tau =6$ in addition to calculations performed on $N_\tau=4$ lattices.
This yields an estimate of finite lattice size effects and allows a 
controlled extrapolation to the continuum limit. 
Moreover, (iii) we use the RHMC algorithm \cite{rhmc} for our calculations.
This eliminates step size errors inherent in earlier studies of
QCD thermodynamics with staggered fermions. 
Without these finite step size errors, a reliable analysis
of finite volume effects is possible since one has excluded the possibility
of finite step size errors and finite volume effects acting in concert.
The RHMC algorithm 
has also been used in other recent studies of QCD thermodynamics with standard 
staggered fermions \cite{aoki,philipsen}.

Our studies of the transition to the high temperature phase of QCD have
been performed on lattices of size $N_\sigma^3 \times N_\tau$ with $N_\tau =
4$ and $6$ and spatial lattice sizes $N_\sigma = 8,~16$, $24$ and $32$. 
We performed calculations for several values of the light to strange quark 
mass ratio, $\hat{m}_l/\hat{m}_s$ for fixed $\hat{m}_s$.  
The strange quark mass has been chosen such that the extrapolation
to physical light quark mass values yields approximately the correct
physical kaon mass value. This led to the choice $\hat{m}_s=0.065$ for our
calculations on $N_\tau=4$ lattices and $\hat{m}_s=0.04$ for the $N_\tau=6$
lattices. Some additional calculations at a larger bare strange quark mass,
$\hat{m}_s=0.1$, have been performed on the $N_\tau=4$ lattices to check the 
sensitivity of our results to the correct choice of the strange quark 
mass. 
Zero temperature calculations have been performed on $16^3\times 32$ lattices. 
On these lattices, hadron masses and the static quark potential have been
calculated. The latter we use to set the scale for the transition temperature,
while the hadron masses specify the physical values of the quark masses.

As will be discussed later in more detail, we use parameters characterizing
the shape of the static quark potential ($r_0$, $r_1$, $\sqrt{\sigma}$)
as well as hadron masses
to set the scale for thermodynamic observables.
At each value of the strange quark mass we have performed
simulations at several light quark mass values corresponding
to a regime of pseudo-scalar (pion) masses\footnote{Here and everywhere
else in this paper we use $r_0=0.469(7)$~fm \cite{gray} to convert
lattice cut-offs to physical units. The $r_0$-parameter is discussed in 
more detail in Section IV.} 
$150~{\rm MeV} \lsim m_{ps} \lsim 500~{\rm MeV}$. A brief overview of
lattice sizes, quark masses and basic simulation parameters used in our 
calculations is given in Table~\ref{tab:parameter}. Further details on
all simulations reported on here and results for some observables are given 
in an Appendix.

The numerical simulation of the QCD partition function has been performed using
the RHMC algorithm \cite{rhmc}. Unlike the hybrid-R algorithm \cite{hybrid}
used in most previous studies of QCD thermodynamics performed with staggered 
fermions, this algorithm has the advantage of being exact, {\it i.e.} 
finite step size errors arising from the discretization of the 
molecular dynamics evolution of gauge fields in configuration space are 
eliminated through an additional Monte Carlo accept/reject step. This is 
possible with the introduction of a rational function approximation for 
roots of fermion determinants appearing in Eq.~\ref{partition}.
We introduce different
step lengths in the integration of gluonic and fermionic parts of 
the force terms that enter the equations of motion for the molecular dynamics
(MD) evolution. During the MD
evolution, we use a 6th order rational approximation for the roots of the
fermion determinants, 
and a more accurate 12th order rational approximation during the Metropolis
accept reject/step.  The choice of these parameters give virtually identical 
results when compared with results obtained using more stringent tolerances.  
We tuned the MD stepsizes to
achieve about (70-80)\% acceptance rate for the new configurations
generated at the end of a MD trajectory of length $\tau_{MD}=0.5$.
As a result of these algorithmic improvements our simulations now run much 
faster compared to the old implementation of the hybrid-R algorithm.  In 
particular, we can use much larger step sizes for our molecular dynamics 
evolution, especially
for the lightest quark masses, resulting in significantly reduced CG counts per
gauge configuration generated.
Details on the tuning of the parameters of the RHMC algorithm used in
our simulations will be given elsewhere \cite{RHMCtuning}.

\begin{table}[t]
\begin{center}
\vspace{0.3cm}
\begin{tabular}{|c|r|r|c|c|c|}
\hline
$N_\tau$ & $\hm_s~~$ & $\hm_l~~$ & $N_\sigma~~$ &\# $\beta$ values & 
max. no. of  conf. \\
\hline
4 & 0.1 & 0.05 & 8 & 10 & 59000\\
  &     & 0.02 & 8 & ~6 & 49000\\
\hline
4 & 0.065 & 0.026 & 8,~16 & 10,~11 & 30000,~ 60000\\
~ & ~     & 0.013 & 8,~16 & ~8,~~7 & 30000,~ 60000\\
~ & ~     & 0.0065 & 8,~16 & ~9,~~6  & 34000,~ 45000 \\
~ & ~     & 0.00325 & 8,~16& ~8,~~5 & 30000,~ 42000 \\
\hline
6 & 0.040 & 0.016 & 16 & 11 & 20000\\
~ & ~     & 0.008 & 16,~32 & ~9,~1 & 62000, 18000\\
~ & ~     & 0.004 & 16,~24 & 7,~6 & 60000, ~8100\\
\hline
\end{tabular}
\end{center}
\caption{Spatial lattice sizes ($N_\sigma$) used for simulations with
different pairs of light and strange quark masses ($\hm_l,\hm_s$) on
lattices with temporal extent $N_\tau$. The fifth column gives the
number of different gauge coupling values at which calculations have
been performed for each parameter set. The last column gives the maximum
number of gauge configurations generated per $\beta$-value.
}
\label{tab:parameter}
\end{table}

\section{Finite temperature simulations}

Our studies of the QCD transition at finite temperature have been
performed on lattices of size $N_\sigma^3\times N_\tau$. The 
lattice spacing, $a$, relates the spatial ($N_\sigma$) and temporal 
($N_\tau$) size of the lattice to the physical volume 
$V= (N_\sigma a)^3$ and temperature $T=1/N_\tau a$, respectively.
The lattice spacing, and thus the temperature, is controlled by the
gauge coupling, $\beta = 6/g^2$, as well as the bare quark masses.

Previous studies of the QCD transition with improved staggered fermions
gave ample evidence that the transition from the low to high
temperature regime of QCD is not a phase transition but rather a 
rapid crossover. The transition is signaled by a 
rapid change in bulk thermodynamic observables (energy density, pressure)
as well as in chiral condensates and the Polyakov loop expectation value,
\begin{eqnarray}
\frac{\langle \bar{\psi}\psi \rangle_q}{T^3} &=& \frac{1}{VT^2}
\frac{\partial \ln Z}{\partial \hat{m}_q}
= \frac{N_\tau^2}{4 N_\sigma^{3}} 
\left\langle {\rm Tr}\; D^{-1}(\hm_q)\right\rangle \;, \; q\; 
\equiv\; l,\; s, \\
\langle L\rangle &=& \left\langle \frac{1}{3N_\sigma^{3}} 
{\rm Tr}\sum_{\bf x} 
\prod_{x_0 = 1}^{N_\tau} U_{(x_0, {\bf x}),\hat{0}} \right\rangle \; , 
\label{orderparameter}
\end{eqnarray}
which are order parameters for a true phase transition in the zero 
and infinite quark mass limit, respectively. 
Note that we have defined the chiral condensate per flavor
degree of freedom, {\it i.e.} the derivative with respect to $\hm_l$ should
be considered as being a derivative with respect to one of the two light
quark degrees of freedom.

On the $N_\tau =4$ lattices we performed calculations at four different
values of the light quark mass, $\hm_l/\hm_s = 0.05$, $0.1$, $0.2$ and
$0.4$ with $\hm_s = 0.065$. This choice of parameters corresponds to
masses of the Goldstone pion ranging from about $150$~MeV to $450$~MeV.
Some additional runs have been performed with a somewhat larger strange
quark mass value, $\hm_s=0.1$, and two values of the light quark mass, 
$\hm_l =0.2\hm_s$ and $0.5\hm_s$, which we used to check the sensitivity of 
our results on the choice of the heavy quark mass (or equivalently the
kaon mass).
On the $N_\tau = 6$ lattices calculations have been performed for three
values of the light quark mass, $\hm_l/\hm_s = 0.1$, $0.2$ and $0.4$ 
with a bare strange quark mass $\hm_s = 0.04$. This covers a range of
pseudo-scalar masses from $240$~MeV to $490$~MeV. 
The  choice of $\hm_s$ insures
that the physical strange quark mass remains approximately constant
for both values of the lattice cut-off. For $N_\tau =4$ we performed 
simulations on lattices with spatial extent $N_\sigma = 8$ and $16$.
For $N_\tau = 6$ most calculations have been performed on $16^3\times 6$
lattices; some checks of finite volume effects have been performed for
$m_l=0.2 m_s$ on a $32^3\times 6$ lattice and for $m_l=0.1 m_s$
on a $24^3\times 6$ lattices.

\begin{figure}[t]
\begin{center}
\begin{minipage}[c]{14.5cm}
\begin{center}
\epsfig{file=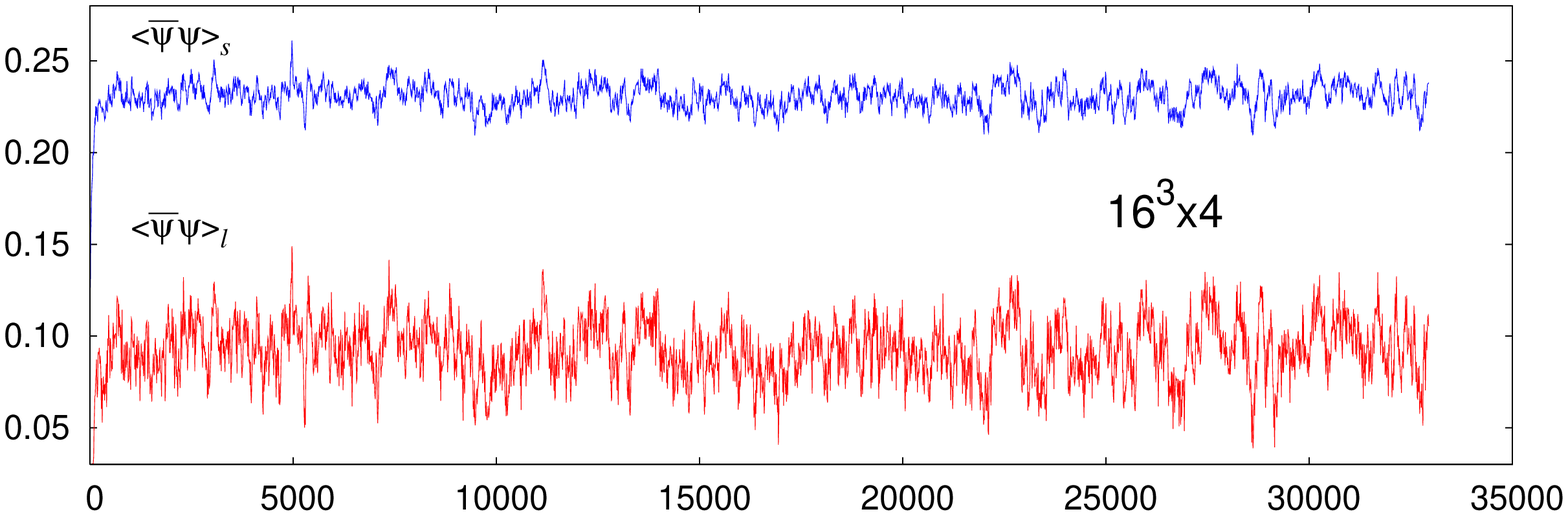, width=12.0cm}
\epsfig{file=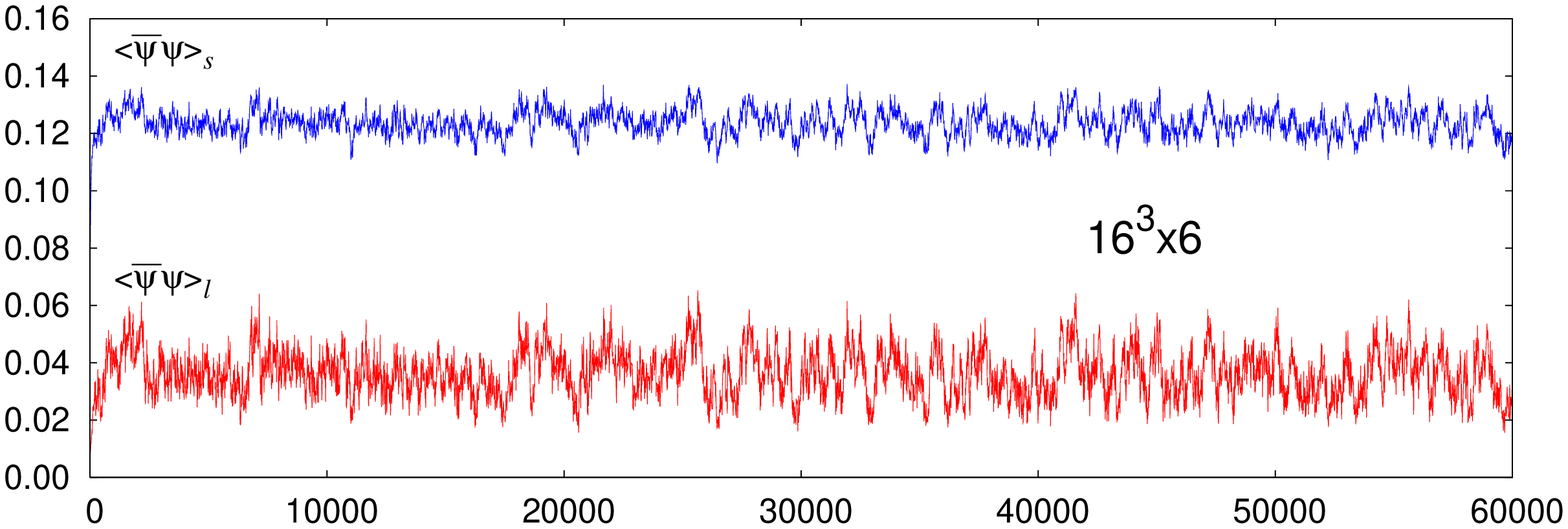, width=12.0cm}
\end{center}
\end{minipage}
\end{center}
\caption{Time history of the 
light and strange quark chiral condensates for the smallest quark masses
used in our simulations on lattices of temporal extent $N_\tau =4$ and $6$
and for values of the gauge coupling in the vicinity of the critical coupling
of the transition on these lattices.
The upper figure shows a run at $\beta= 3.305$ with $\hm_l=0.05 \hm_s$ 
on a $16^3\times 4$ lattice and the lower figure is for $\beta=3.46$ 
and $\hm_l=0.1 \hm_s$ on a $16^3\times 6$ lattice.
}
\label{fig:history}
\end{figure}

For each parameter set $(\beta,\; \hm_l,\; \hm_s)$ we generally generated 
more than 10000, and in some cases up to 60000, gauge field configurations.
While the Polyakov loop expectation value and its susceptibility have 
been calculated on each gauge field configuration, the chiral condensates
and their susceptibilities have been analyzed only on every $10^{th}$ 
configuration using unbiased noisy estimators with $10$ noise vectors
per configuration. We have monitored the auto-correlation times in all 
our runs. From correlation functions of the gauge action we typically find 
auto-correlation times $\tau_{MD}$ of ${\cal O} (100)$ configurations.
They can rise up to ${\cal O} (250)$ configurations in the vicinity of
the transition temperature. Our data samples thus typically 
contain a few hundred statistically independent configurations for each
parameter set. We show two time histories of chiral condensates in the
transition region in Figure~\ref{fig:history}.
All simulation parameters, results on auto-correlation times, the light and 
heavy quark condensates, the Polyakov loop expectation value, and the 
corresponding susceptibilities are summarized in Tables~A.1 to A.9  which are 
presented in the Appendix.

\begin{figure}[t]
\begin{center}
\begin{minipage}[c]{14.5cm}
\begin{center}
\epsfig{file=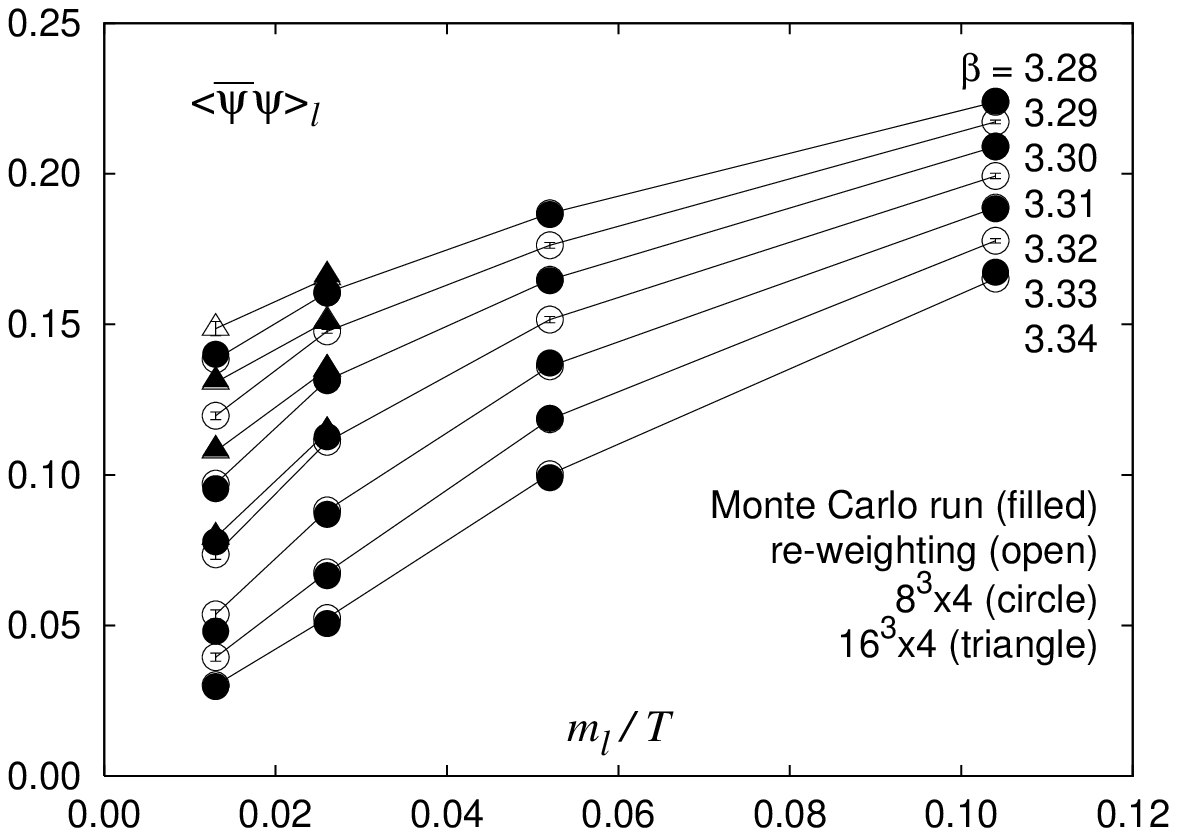, width=7.0cm}
\epsfig{file=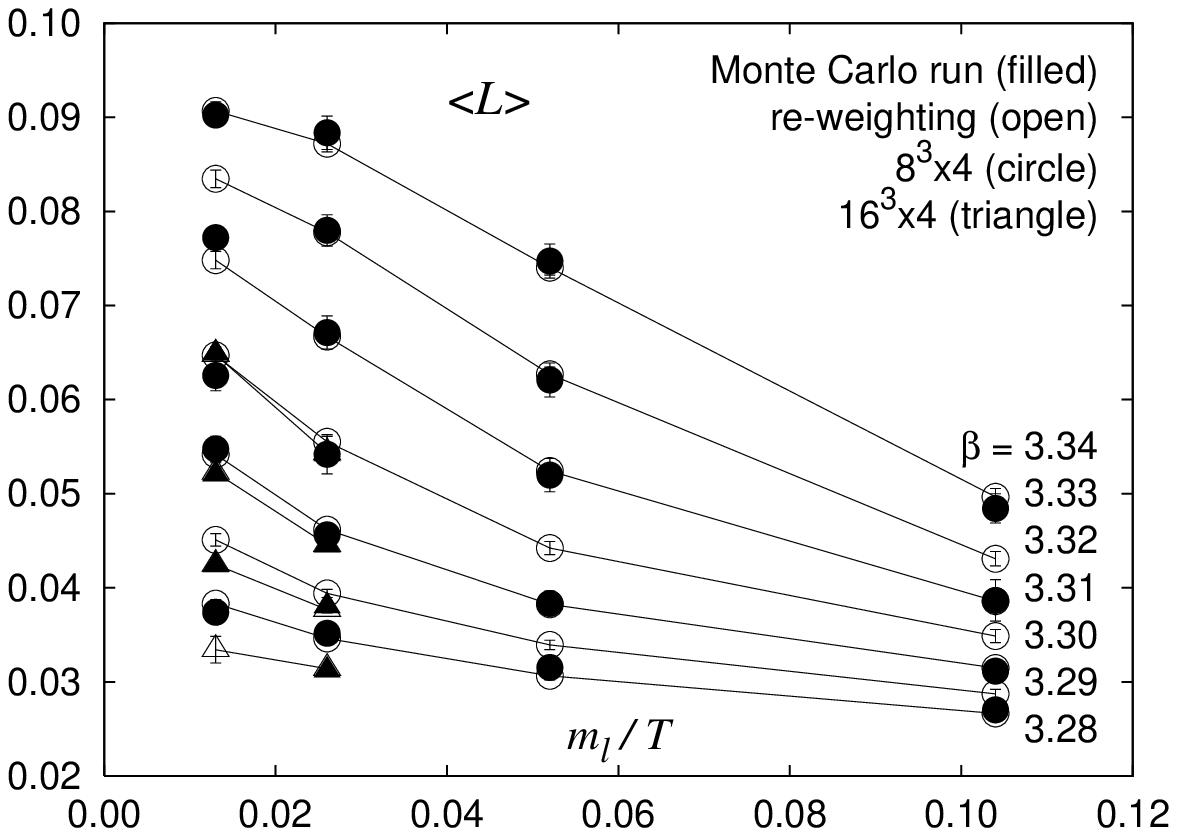, width=7.0cm}
\end{center}
\end{minipage}
\end{center}
\caption{The light quark chiral condensate in units of $a^{-3}$ (left) and 
the Polyakov loop expectation value (right) as function of the bare light 
quark mass in units of the temperature, $m_l/T\equiv \hm_l N_\tau$ for 
fixed $\beta$ and $\hm_s=0.065$ on lattices of size 
$8^3\times 4$ (circle) and $16^3\times 4$ (triangles). Shown are results
for various values of $\beta$ ranging from $\beta = 3.28$ to $\beta = 3.4$
(top to bottom for $\langle \bar{\psi}\psi\rangle$ and bottom to top for 
$\langle L \rangle$). Full and open symbols show results obtained from 
direct simulations and Ferrenberg-Swendsen interpolations, respectively.
}
\label{fig:pbp_4}
\end{figure}

In Figure~\ref{fig:pbp_4}(left) we compare results for the light quark
chiral condensate calculated on lattices of size $8^3\times 4$
and $16^3\times 4$. It clearly reflects the presence of finite volume
effects at small values of the quark mass. While finite volume effects seem to be  
negligible for $\hm_l/\hm_s \ge 0.2$, for $\hm_l/\hm_s =0.1$ we observe 
a small but statistically significant volume dependence for the chiral condensate
as well as for the Polyakov loop expectation value. This volume dependence is even 
more pronounced for $\hm_l/\hm_s =0.05$ and seems to be stronger at low 
temperatures.  While the value of the chiral condensate
increases with increasing volume the Polyakov loop expectation value decreases
(Figure~\ref{fig:pbp_4}(right)).

In a theory with Goldstone bosons, e.g. in $O(N)$-symmetric spin models, it is
expected that in the broken phase the order parameter, ${\cal O}$, changes with 
the symmetry breaking field, $h$, as ${\cal O}(h)-{\cal O}(0)\sim h^{1/2}$ 
\cite{wallace}.  This behavior has also been found in QCD with adjoint quarks, 
{\it i.e.} $\langle \bar{\psi}\psi\rangle \sim c_0\; +\; c_1 (m_l/T)^{1/2}$ 
\cite{lutgemeier,engels}. Our current analysis of the quark mass dependence
of the chiral condensate is not yet accurate enough and has not yet been 
performed at small enough quark masses to verify this behavior explicitly. 
We will analyze the light quark mass limit in more detail elsewhere. 

We use  the Polyakov loop susceptibility as well as the disconnected part 
of the chiral susceptibility to locate the transition temperature to the 
high temperature phase of QCD,
\begin{eqnarray}
\chi_L &\equiv& N_\sigma^{3}  \left( 
\langle  L^2 \rangle - \langle L \rangle^2 \right) \; ,
\label{sus_L}\\
\frac{\chi_q}{T^2} &\equiv& \frac{N_\tau}{16N_\sigma^{3}} \left( 
\left\langle \left( {\rm Tr}\; D^{-1}(\hm_q)\right)^2 \right\rangle -
\left\langle {\rm Tr\;} D^{-1}(\hm_q)\right\rangle^2\right) \;, \; 
q\; \equiv\; l,\; s . \label{sus_chi}
\label{sus_m}
\end{eqnarray}

\begin{figure}[t]
\begin{center}
\begin{minipage}[c]{14.5cm}
\begin{center}
\epsfig{file=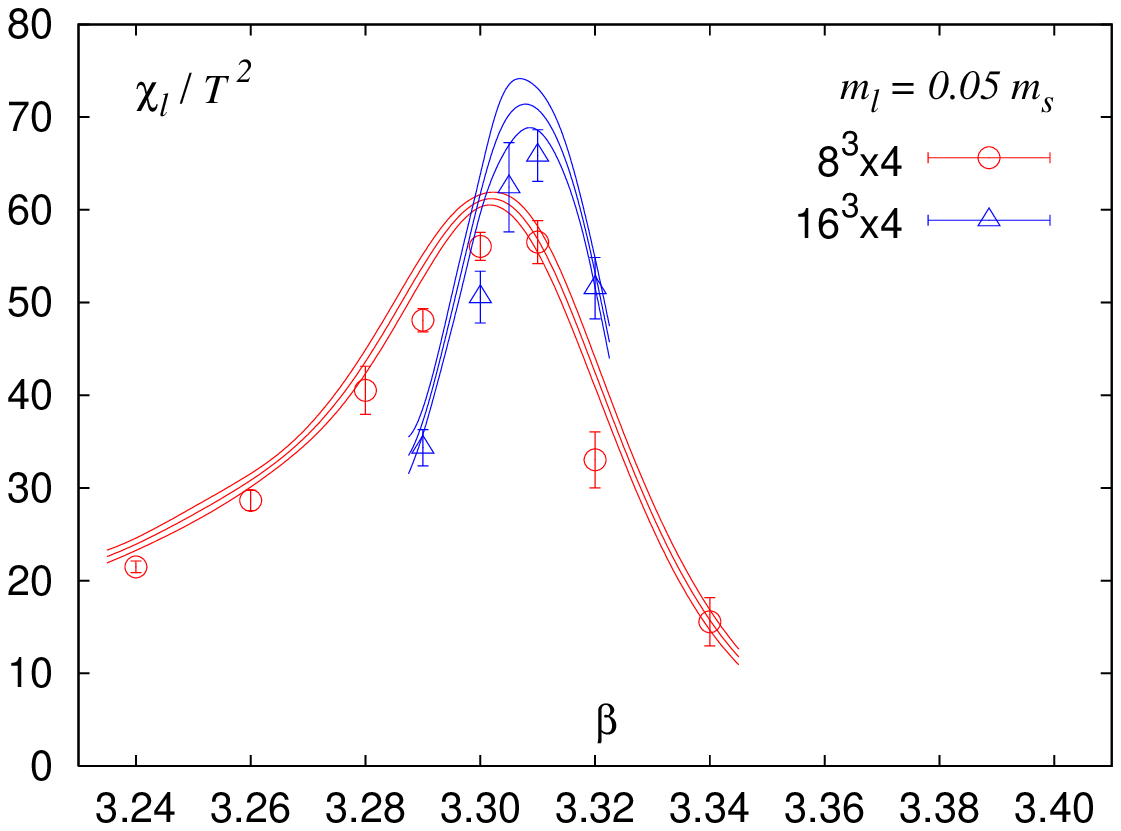, width=7.0cm}
\epsfig{file=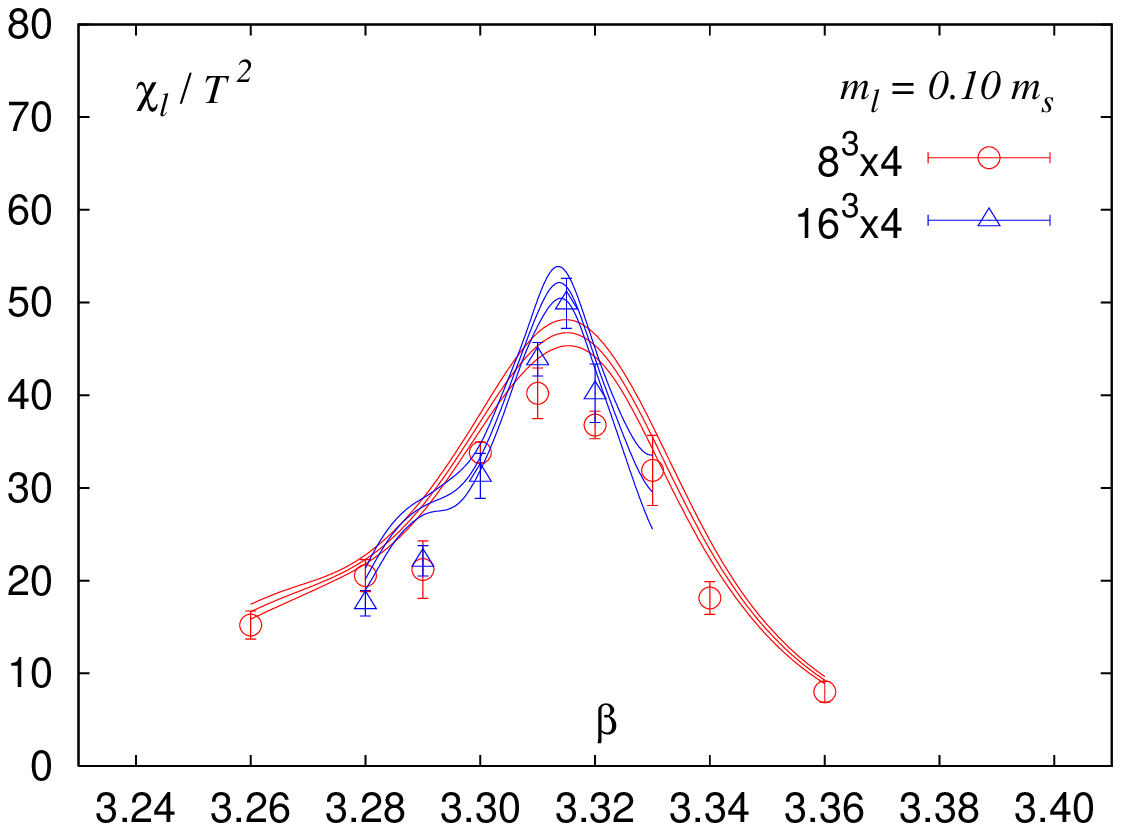, width=7.0cm}
\epsfig{file=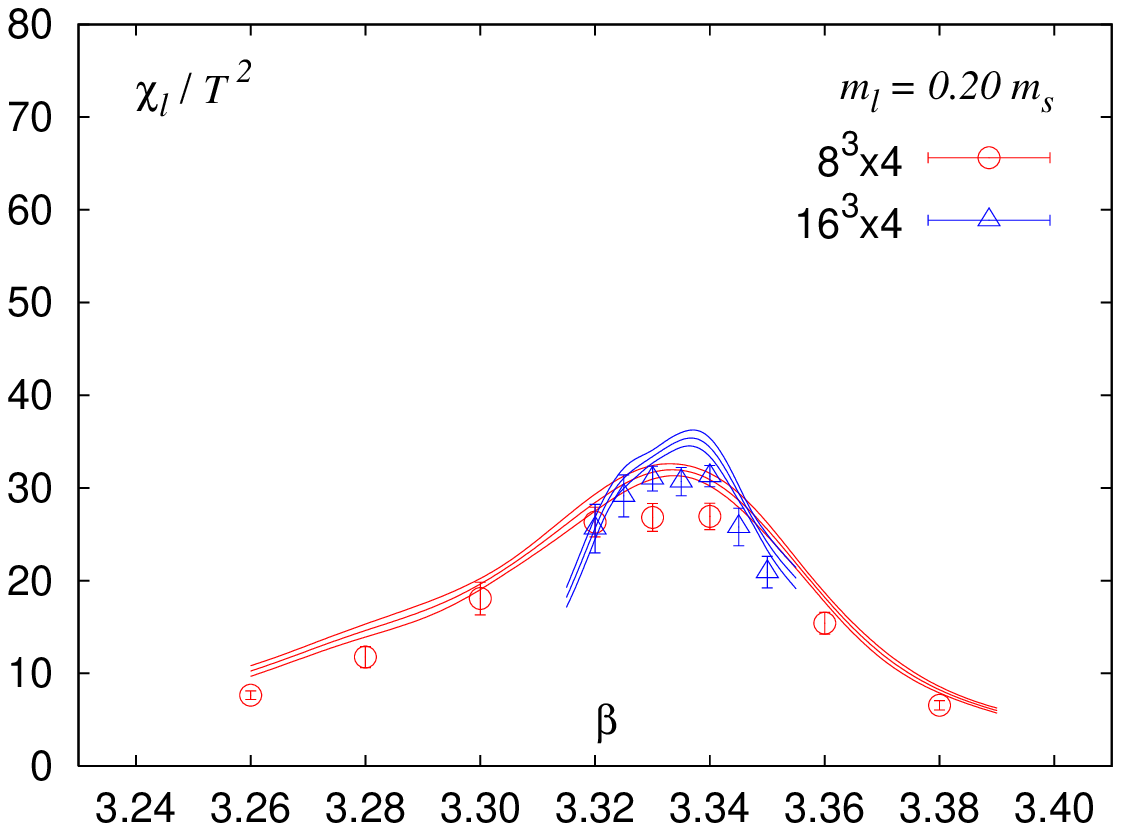, width=7.0cm}
\epsfig{file=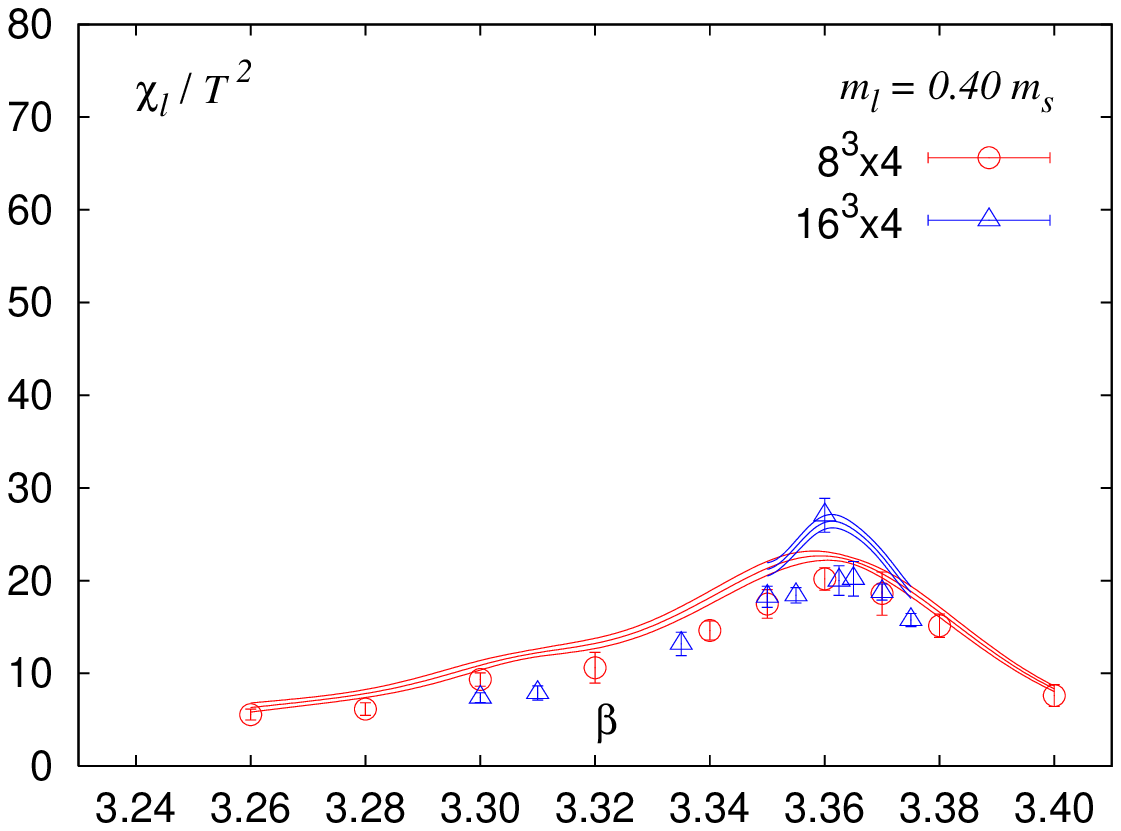, width=7.0cm}
\end{center}
\end{minipage}
\end{center}
\caption{The disconnected part of the light quark chiral susceptibility on 
lattices of size
$8^3\times 4$ (squares) and $16^3\times 4$ (circles) for four different
values of the light quark mass. The curves show  Ferrenberg-Swendsen 
interpolations of the data points obtained from multi-parameter histograms
with an error band coming from Ferrenberg-Swendsen reweightings performed 
on different jackknife samples.} 
\label{fig:chi_4}
\end{figure}

\begin{figure}[t]
\begin{center}
\begin{minipage}[c]{14.5cm}
\begin{center}
\epsfig{file=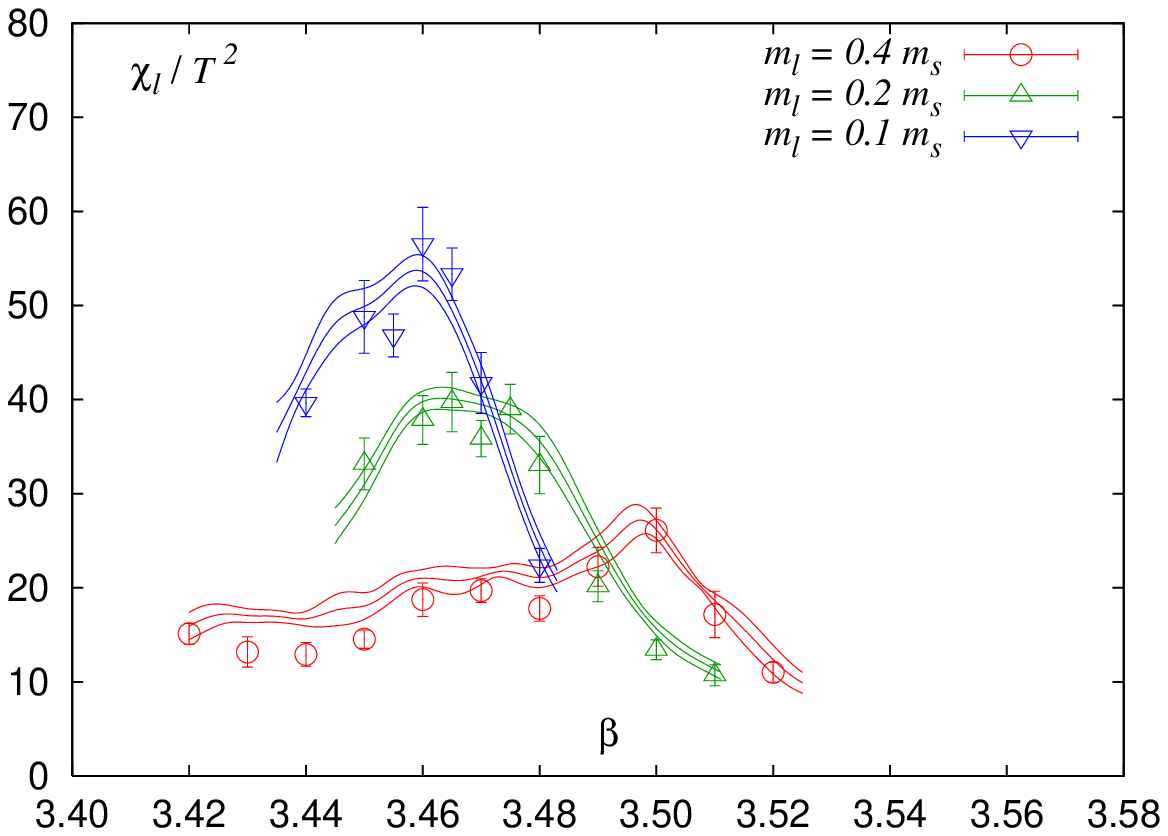, width=7.0cm}
\epsfig{file=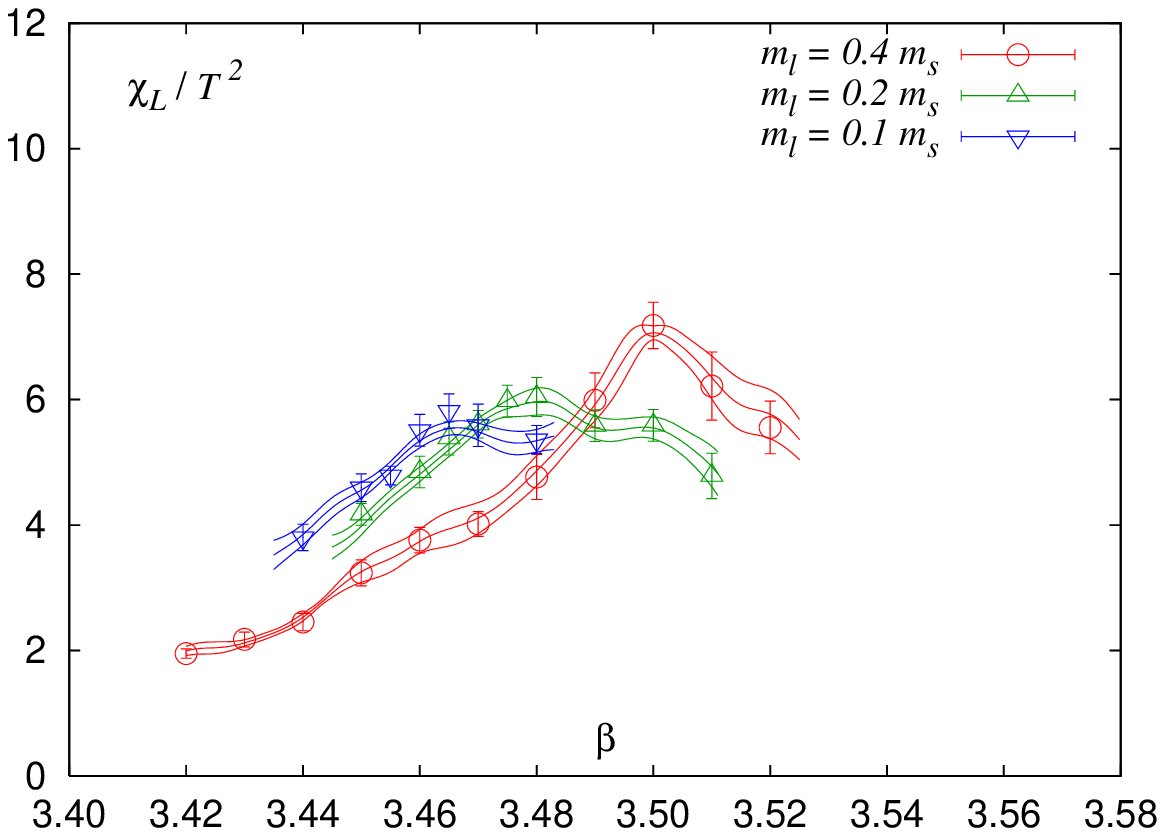, width=7.0cm}
\end{center}
\end{minipage}
\end{center}
\caption{The disconnected part of the light quark chiral susceptibility (left)
and the Polyakov loop susceptibility (right) on lattices of size
$16^3\times 6$ for three different values of the light quark mass. Curves 
show Ferrenberg-Swendsen interpolations as discussed in the caption of Fig.~2}
\label{fig:chi_6}
\end{figure}

In Figure~\ref{fig:chi_4} we show results for the disconnected part of the
light quark chiral susceptibility, $\chi_l$, calculated on $8^3\times 4$ 
and $16^3\times 4$ lattices. Results for  $\chi_l$ and the Polyakov loop
susceptibility, $\chi_L$, obtained from our calculations on $16^3\times 6$ 
lattices are 
shown in Figure~\ref{fig:chi_6}. The location of peaks in the susceptibilities
has been determined from a Ferrenberg-Swendsen reweighting of data in the 
vicinity of the peaks. Errors on the critical couplings determined in this
way have been obtained from a jackknife analysis where Ferrenberg-Swendsen
interpolations have been performed on different sub-samples.
In agreement with earlier calculations we find that the position of  
peaks in $\chi_l$ and $\chi_L$ show only little volume dependence and that 
the peak height changes only little, although the maxima become somewhat more 
pronounced on the larger lattices. This is consistent with the transition being
a crossover rather than a true phase transition in the infinite volume limit.

\begin{figure}[t]
\begin{center}
\begin{minipage}[c]{14.5cm}
\begin{center}
\epsfig{file=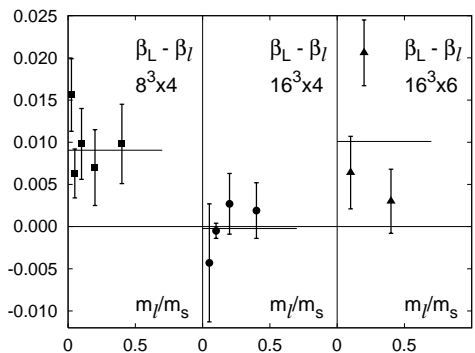, width=7.0cm}
\end{center}
\end{minipage}
\end{center}
\caption{The difference of gauge couplings at the location of 
peaks in the Polyakov loop and chiral susceptibilities, 
$\beta_{c,L}-\beta_{c,l}$. Shown are results from calculations on
$8^3\times 4$ (left), $16^3\times 4$ (middle) and $16^3\times 6$ 
(right).}
\label{fig:diff}
\end{figure}

Although differences in the critical coupling extracted from $\chi_L$ and
$\chi_l$ are small we find that on small lattices the peak in the Polyakov loop 
susceptibility is located at a systematically larger value of the gauge 
coupling $\beta$. In a finite volume this is, of course, not
unexpected, and in the infinite volume limit an ambiguity in identifying the
transition point may also remain for a crossover transition. Nonetheless, we 
observe that the difference  $\beta_{c,L}-\beta_{c,l}$ decreases with increasing 
volume and is within errors consistent with zero for $16^3\times 4$, which has
the largest spatial volume expressed in units of the temperature,
$TV^{1/3}=4$. On the smallest lattice, $8^3\times 4$, we find
$\beta_{c,L}-\beta_{c,l}\simeq 0.0077(9)$. Within the statistical accuracy
of our data we also do not find any systematic quark mass dependence of this 
difference, $\beta_{c,L}-\beta_{c,l}$, which is shown in Figure~\ref{fig:diff}
for the 3 different system sizes used in our calculations. 

The peak positions, $\beta_c(\hm_l,\hm_s, N_\tau)$, in the chiral and 
Polyakov loop
susceptibilities are generally well determined. An exception is our data set
for $\hm_l/\hm_s=0.2$ on the $16^3\times 6$ lattice which shows a quite broad 
peak in $\chi_l$. Consequently we find here the largest difference 
$\beta_{c,L}-\beta_{c,l}\simeq 0.019(7)$ which also has the largest
statistical error. 

The error on $\beta_{c,(L,l)}(\hm_l,\hm_s, N_\tau)$ translates, of course, into 
an uncertainty for the lattice spacing $a(\beta_c)$ which in turn contributes 
to the error on the transition temperature. In order to get a feeling for the 
accuracy required in the determination of $\beta_c$ 
we give here an estimate for the dependence of the lattice cut-off
on the gauge coupling, which will be determined and discussed in more detail in 
the next section. From an analysis of scales related to the static quark potential
at zero temperature we deduce that, in the regime of couplings relevant for
our finite temperature calculations, a shift in the gauge coupling by 
$\Delta\beta =0.02$ corresponds to a change in the lattice cut-off of
about 5\%. An uncertainty in the determination of the critical coupling of
about $0.01$ thus translates into a 2.5\% error on $T_c$. 

We summarize our results for the critical couplings determined from
peaks in $\chi_L$ and $\chi_l$, respectively, in Table~\ref{tab:betac}. 
As the peak positions in both quantities apparently differ systematically 
on finite lattices and for finite values of the quark mass, we use the average 
of both values as an estimate for the critical coupling, $\beta_c$, for the 
transition to the high temperature phase of QCD. We include the 
deviation of $\beta_{c,L}$ and $\beta_{c,l}$ from this average value as a 
systematic error in $\beta_c$ and add it quadratically to the statistical
error. These averaged critical couplings are given in the last column of 
Table~\ref{tab:betac}. Even with this conservative error estimate the 
uncertainty in the determination of the critical coupling is in almost all
cases smaller than $0.01$, {\it i.e.} the uncertainty in the
determination of $\beta_c$ will amount to about 2\% error in the  
determination of $T_c$.

\begin{table}[t]
\begin{center}
\vspace{0.3cm}
\begin{tabular}{|c|l|l|r|c|c|c|c|}
\hline
$N_\tau$&$\hm_s$ & $~~\hm_l$& $N_\sigma$ & $\beta_{c,L}$ [from $\chi_L$]
& $\beta_{c,l}$ [from $\chi_l$] & $\beta_{c,s}$ [from $\chi_s$] 
& $\beta_c$ [averaged] \\
\hline
4&0.1&0.05000& 8&   3.4248(46)&   3.4125(32)&   3.4132(36)&   3.4187( 83) \\
~&~&0.02000& 8&   3.3740( 31)&   3.3676( 18)&   3.3723( 35)&   3.3708( 48) \\
\hline
4&0.065&0.02600&16&   3.3637(20)&   3.3618(12)&   3.3615(13)&   3.3627( 25) \\
~&~&0.02600& 8&   3.3661(25)&   3.3563(21)&   3.3588(28)&   3.3612( 59) \\
~&~&0.01300&16&   3.3389(18)&   3.3362(17)&   3.3374(17)&   3.3376( 28) \\
~&~&0.01300& 8&   3.3419(23)&   3.3349(21)&   3.3398(31)&   3.3384( 47) \\
~&~&0.00650&16&   3.3140(~5)&   3.3145(~3)&   3.3132(~4)&   3.3143( ~6) \\
~&~&0.00650& 8&   3.3239(30)&   3.3141(11)&   3.3198(24)&   3.3190( 58) \\
~&~&0.00325&16&   3.3042(43)&   3.3085(26)&   3.3132(69)&   3.3064( 55) \\
~&~&0.00325& 8&   3.3087(18)&   3.3024(10)&   3.3047(17)&   3.3056( 37) \\
\hline
6&0.04&0.01600&16&   3.5002(25)&   3.4973(12)&   3.4967(12)&   3.4988( 32) \\
~&~&0.00800&16&   3.4801(19)&   3.4595(19)&   3.4614(44)&   3.4698(106) \\
~&~&0.00400&16&   3.4668(24)&   3.4604(18)&   3.4603(11)&   3.4636( 44) \\
\hline
\end{tabular}
\end{center}
\caption{Critical couplings determined from the location of peaks in the 
Polyakov loop susceptibility as well as in the disconnected parts of the 
light and strange quark chiral susceptibilities. The last column gives the
average of $\beta_{c,L}$ and $\beta_{c,l}$.}
\label{tab:betac}
\end{table}

In addition to the light quark condensate and its susceptibility we also
have analyzed the strange quark condensate and its susceptibility, $\chi_s$.
We find that the light and heavy quark condensates are strongly correlated,
which is easily seen in the MD-time evolution of these quantities. Already
on the smallest lattices
the position of the peak in the heavy quark susceptibilities is consistent with 
that deduced from the light quark condensate. On the larger, $N_\sigma=16$, 
lattices the difference  $|\beta_{c,l}-\beta_{c,s}|$ is in all cases zero
within statistical errors, which are about $3\cdot 10^{-3}$. Any 
temperature difference in the crossover behavior for the light and strange
quark sector of QCD, which sometimes is discussed in phenomenological 
models, thus is below the $1$~MeV level.

\section{Zero Temperature Scales}

In order to calculate the transition temperature in terms of an 
observable that is experimentally accessible and can be used to 
set the scale for $T_c$ we have to perform a zero temperature 
calculation at the critical couplings $\beta_c$ determined
in the previous section. This will allow us to eliminate the unknown 
lattice cut-off, $a(\beta_c)$, which determines  $T_c$  on a lattice with
temporal extent $N_\tau$, {\it i.e.} $T_c= 1/N_\tau a(\beta_c)$. 
To do so we have performed calculations at zero temperature, {\it i.e.}
on lattices of size $16^3\times 32$, and calculated several hadron
masses as well as the static quark potential. From the latter we determine 
the string tension and 
extract short distance scale parameters $r_0,~r_1$, which are 
defined as separations between the static quark anti-quark sources at which the 
force between them attains certain values \cite{Sommer},
\begin{equation}
r^2\frac{{\rm d} V_{\bar{q}q}(r)}{{\rm d}r}\biggl|_{r=r_0} = 1.65   \;\; , \;\;
r^2\frac{{\rm d} V_{\bar{q}q}(r)}{{\rm d}r}\biggl|_{r=r_1} = 1.0   \; .
\label{r0}
\end{equation}
Although these scale parameters are not directly accessible to experiment they 
can be well estimated from heavy quarkonium phenomenology. Moreover, 
they have been determined quite accurately in lattice calculations
through  a combined analysis
of the static quark potential \cite{MILCpotential} and level splittings
in bottomonium spectra \cite{gray}. Both these calculations have been performed
on identical sets of gauge field configurations. We will use the value for
$r_0$ determined in the bottomonium calculation \cite{gray} 
for all conversions of lattice results to physical units,
\begin{equation}
r_0 = 0.469(7)\;{\rm fm}  \; .
\label{r0_fm}
\end{equation}

Our zero temperature calculations have been performed at values of the
gauge coupling in the vicinity of 
the $\beta_c$ values listed in the last column of Table~\ref{tab:betac}.
We typically generated several 
thousand configurations and analyzed the hadron spectrum and static 
quark potential on every $10^{th}$ configuration. A summary of our
zero temperature simulation parameters is given in Table~\ref{tab:scale} 
together with the two scales characterizing the static quark potential,
$r_0/a$ and $\sqrt{\sigma}a$, expressed in lattice units. These scales have 
been obtained by using the simple Cornell form to fit our numerical
results for the static quark potential, 
$V_{\bar{q}q}(r) = -\alpha/r + \sigma r + c$.
With this fit-ansatz, which does not include a possible running of the
coupling $\alpha$, the force entering the definition of $r_0$ is easily
calculated and we find from Eq.~\ref{r0}, 
$r_0 \equiv \sqrt{(1.65 - \alpha)/\sigma}$. More details on the analysis
of the static quark potential and the precise form for the fit ansatz used
by us will be given in the next subsection.

In addition we also have calculated some meson masses, the mass of the lightest
pseudo-scalar in the light quark sector, $m_{ps}$ and the strange quark
sector, $m_{s\bar{s}}$; and the pseudo-scalar heavy-light meson, $m_K$. 
Results for these masses are also given in Table~\ref{tab:scale}. 
They have been obtained from point-wall correlation functions using 
a $Z_2$-wall source. The correlation functions have been fitted to a double 
exponential ansatz that takes into account the two lowest states contributing
to the staggered fermion correlation functions. We have varied the lower limit,
$r_{min}$, of the fit range to check for the stability of our fits. For
the masses displayed in Table~\ref{tab:scale} 
stable results typically are found for $r_{min}\gsim 6$ for $N_\tau = 4$
and $r_{min}\gsim 8$ for $N_\tau = 6$. 
In the following we discuss in more
detail our analysis of the static quark potential. 

\begin{table}[t]
\begin{center}
\vspace{0.3cm}
\begin{tabular}{|l|l|c|c|c|c|c|c|c|c|}
\hline
$\hm_s$ & $~~\hm_l$ & $~~\beta$ &\# conf.& $m_{ps}a$ & $m_{s\bar{s}}a$ & $m_Ka$ & $r_0/a$
&$(r_0/a)_{smooth}$&$\sqrt{\sigma}a$\\
\hline
0.1&0.05&3.409&600&0.7075(3)&0.9817(2)&0.8571(3)& 2.0525(36)(89)&-& 0.5564(17)(79) \\
~&0.02&3.371&560&0.4583(2)&0.9854(4)&0.7748(3)& 2.0178(45)(56)&2.0097& 0.5651(24)(92)\\
\hline
0.065&0.026&3.362&500&0.5202(4)&0.8045(3)&0.6794(6)& 2.0250(59)(75)&2.0337& 0.5580(26)(89)\\
~&0.013&3.335&400&0.3733(3)&0.8072(4)&0.6339(4)& 1.9801(47)(11)&1.9803& 0.5675(24)(90)\\
~&0.0065&3.31&750&0.2656(4)&0.8089(2)&0.6092(5)& 1.9047(40)(132)&1.9018& 0.5910(25)(116)\\
~&0.00325&3.30&400&0.1888(6)&0.8099(3)&0.5948(3)& 1.8915(59)(136)&1.8750& 0.5888(34)(95)\\
\hline
0.04&0.016&3.50&294&0.3864(6)&0.5988(6)&0.5048(6)& 3.0061(143)(92)&3.0136& 0.3766(27)(33)\\
~&0.008&3.47&500&0.2831(13)&0.6097(6)&0.4789(7)& 2.8953(96)(56)&2.8736& 0.3867(19)(40)\\
~&0.004&3.455&410&0.2043(10)&0.6143(6)&0.4634(7)& 2.8030(75)(51)&2.8056& 0.4016(19)(43)\\
\hline
\end{tabular}
\end{center}
\caption{Simulation parameter for the scale setting runs on $16^3\times 32$ lattices
and results obtained for light and heavy quark pseudo-scalars ($m_{ps}$ and 
$m_{s\bar{s}}$), 
the kaon mass and scale parameters of the heavy quark potential. In column 4 we give the number of 
configurations actually used in the analysis. Column 9 show the {\it smoothed} values
for $r_0/a$ obtained from the fit ansatz given in Eq.\ref{interpolate}. Results for the largest 
quark mass pair, $(\hm_s,\hm_l)=(0.1,0.05)$, have not been included in the fit.
}
\label{tab:scale}
\end{table}

\subsection{The static quark potential}
 
The static quark potential at fixed spatial separation has been obtained
from an extrapolation of ratios of Wilson loops to infinite time
separation. The spatial transporters
in the Wilson loop were constructed from spatially smeared links
which have been obtained iteratively by adding space-like 
3-link staples with a relative weight $\gamma=0.4$ to the links and
projecting this sum back to an element of the $SU(3)$ gauge 
group (APE smearing). This
process has been repeated 10 times. We have calculated
the potential for on-axis as well as off-axis spatial separations. 
As we have to work on still rather coarse
lattices and need to know the static quark potential at
rather short distances (in lattice units) we have to deal with
violations of rotational symmetry in the potential. 
In our analysis of the potential we take care of this by adopting a strategy 
used successfully in the analysis of static quark potentials \cite{Sommer}
and heavy quark free energies \cite{zantow}. 
We replace the Euclidean distance on the lattice,
$(r/a)^2 = n_x^2+n_y^2+n_z^2$, by $r_I/a$ which
relates the separation between the static
quark and anti-quark sources to the Fourier transform of the tree-level
lattice gluon propagator, $D_{\mu \nu}$, {\it i.e.}
\begin{equation}
(r_I/a)^{-1}=4 \pi \int_{-\pi}^{\pi} \frac{d^3 k}{{(2 \pi)}^3}
\exp(i \vec{k}\cdot \vec{n}) D_{00}(k) \quad .
\end{equation}
which defines the lattice Coulomb potential.
Here the integers $n=(n_x,n_y,n_z)$ label the spatial components of the
4-vector for all lattice sites and $D_{00}$ is the time-like component of
$D_{\mu \nu}$. For the ${\cal O}(a^2)$ improved gauge action used here 
this is given by
\begin{equation}
D^{-1}_{00}(k)=
4 \sum_{i=1}^3 \left( \sin^2\frac{k_i}{2}+\frac{1}{3} \sin^4\frac{k_i}{2}
\right) \quad .
\end{equation}
This procedure removes most of the short distance lattice artifacts.
It allows us to perform fits to the heavy quark potential with the
3-parameter ansatz, 
\begin{equation}
V_{\bar{q}q}(r)= -\frac{\alpha}{r_I} + \sigma r_I +c \; .
\label{V3}
\end{equation}
Fit results for $\sqrt{\sigma}a$ and  
$r_0/a=\sqrt{(1.65-\alpha)/\sigma a^2}$ obtained with this ansatz
are given in Table~\ref{tab:scale}. Errors on both quantities have been 
calculated from a jackknife analysis. We also performed fits with a 
4-parameter ansatz commonly used in the literature,
\begin{equation}
V_{\bar{q}q}(r)= -\frac{\alpha}{r} + \sigma r +\alpha'\left(\frac{1}{r_I} - 
\frac{1}{r} \right) + c \; .
\label{V4}
\end{equation}
Using this ansatz for our fits, we generally obtain results which are compatible 
with the fit parameters extracted from the 3-parameter fit.
We combine the difference between the 4-parameter fit result and the 
3-parameter fit with differences that arise when changing the fit range for 
the potentials and quote this as a systematic error. 
This is given as a second error for $r_0/a$ and $\sqrt{\sigma}a$ listed in
Table~\ref{tab:scale}. Using Eq.~\ref{r0_fm} we find that the
lattice spacings corresponding to the relevant coupling range explored in 
our $N_\tau=4$ and $6$ calculations correspond to $a\simeq 0.24$~fm and
$a\simeq 0.17$~fm, respectively.  
As can be seen from Table~\ref{tab:scale} we obtain values for $r_0\sqrt{\sigma}$
between $1.11$ and $1.13$. These values are about 2\% larger
than those obtained on finer lattices by the MILC collaboration 
\cite{MILCpotential}. 

We have determined the scale parameter $r_0$ in units of the lattice
spacing for 9 different parameter sets $(\hm_l, \hm_s, \beta)$. This 
allows to interpolate between different values of the gauge coupling
and quark masses.
We use a renormalization group inspired ansatz \cite{allton} which takes 
into account the quark mass dependence of $r_0/a$ 
\cite{Bernard04} and which approaches, in the weak coupling limit,
the 2-loop  $\beta$-function for three massless flavors,
\begin{equation}
(r_0/a)^{-1} = 
R(\beta) (1 +B \hat{a}^2(\beta) + C \hat{a}^4(\beta))  
{\rm e}^{A (2 \hm_l + \hm_s)+D} 
\; .
\label{interpolate}
\end{equation}
Here $R(\beta)$ denotes the 2-loop $\beta$-function  
and $\hat{a}(\beta) =R(\beta)/R(\bar{\beta})$ with $\bar{\beta}=3.4$ chosen
as an arbitrary normalization point.
A fit to 14 values for $r_0/a$, which include 8 of the 9 values for 
$r_0/a$ given in Table~\ref{tab:scale} and additional data obtained in
our studies of 3-flavor QCD \cite{RHMCtuning}, gives $A=1.45(5)$, $B=1.20(17)$, 
$C=-0.21(6)$ and $D= 2.41(5)$ with a $\chi^2/dof = 0.9$.
We use this interpolation formula to set the scale for the transition
temperature.

\subsection{The physical point}

Our goal is to determine the transition temperature at the physical point,
{\it i.e.} for quark masses that correspond to the physical light and
strange quark masses that reproduce the experimentally known hadron
mass spectrum at zero temperature. To do so we reduce the bare light quark
mass, $\hm_l$, keeping $\hm_s$ fixed to an appropriate value that yields the 
physical value for the pion mass expressed, for instance, in units of $r_0$, 
{\it i.e.} $m_{ps}r_0 = 0.321(5)$. The strange quark mass should also be chosen 
such that, at this point, one of the strange meson masses is reproduced. 
For this purpose we monitor the value of the kaon mass and the strange 
pseudo-scalar\footnote{The mass of the strange pseudo-scalar may be
estimated as $m_{s\bar{s}}=\sqrt{2 m^2_K-m^2_\pi} = 686$~MeV \cite{MILCmasses}.}, 
$m_{s\bar{s}}$.  
In the continuum limit the physical point is then given as
$(m_{ps}r_0, m_K r_0)= (0.321(5), 1.177(18))$ where the error
reflects the uncertainty in $r_0$ \cite{gray}. At this point the strange 
pseudo-scalar in units of $r_0$ is given by $m_{s\bar{s}} r_0 = 1.631(24)$. 

\begin{figure}[tb]
\begin{center}
\begin{minipage}[c]{14.5cm}
\begin{center}
\epsfig{file=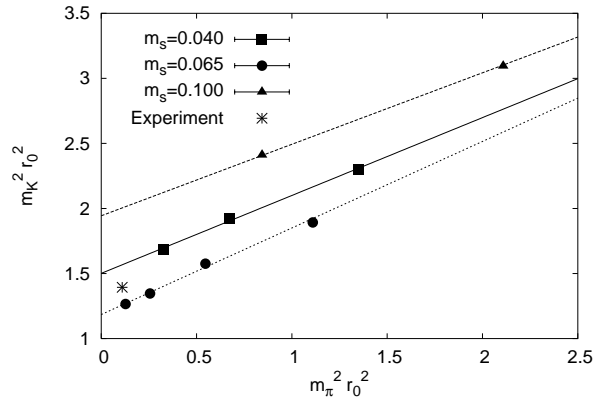, width=8.0cm}
\end{center}
\end{minipage}
\end{center}
\caption{The square of the kaon mass in units of $r_0^2$ versus the
square of the pseudo-scalar meson mass also expressed in units of $r_0^2$.
Shown are results for the three different sets of bare strange quark
masses, $\hm_s=0.04$, (squares) $0.065$ (circles) and $0.1$ (triangles). 
The star shows the location of the physical point using $r_0=0.469$~fm.
The values of the light quark
masses and gauge couplings at which the zero temperature calculations 
on $16^3\times 32$ lattice have been performed can be found in 
Table~\ref{tab:scale}.
}
\label{fig:kaon}
\end{figure}

In Figure~\ref{fig:kaon}, we show the kaon masses in units of $r_0$, 
corresponding to the different sets of light
and heavy quark mass values used in our calculations,  
plotted versus pseudo-scalar masses in units of $r_0$. These data
are also given in Table~\ref{tab:scale}.
It can be seen that the two bare strange quark mass values, $\hm_s=0.065$ 
and $0.04$, used in our finite temperature calculations on $N_\tau=4$ 
and $6$ lattices, respectively, allow us to approach the physical point
in the light quark mass limit. For $m_{ps}r_0=0.321$ we obtain 
$m_Kr_0 = 1.12$ and $1.25$ for the two parameter sets, which
agrees with the continuum value for the kaon mass within 6\%.
The strange pseudo-scalar mass is, as expected, almost independent of the 
value of the light quark mass. Using the data displayed in 
Table~\ref{tab:scale} we find from
a linear extrapolation to the physical point, $m_{s\bar{s}}r_0 = 1.53(2)$
($\hm_s =0.065$ data set) and $m_{s\bar{s}}r_0 = 1.69(2)$ ($\hm_s =0.04$ 
data set), respectively. This too agrees with the continuum value within 6\%.

For the third parameter set, $\hm_s=0.1$, we obtain extrapolated values
for the kaon mass, $m_K r_0 = 1.41$ and for the strange pseudo-scalar
$m_{s\bar{s}}r_0 =1.96$, which both are about 20\% larger than the
physical values.
We use this parameter set to verify the insensitivity of $T_c r_0$ to the
precise choice of the strange quark mass. 

\section{The transition temperature}

\begin{figure}[tb]
\begin{center}
\begin{minipage}[c]{14.5cm}
\begin{center}
\epsfig{file=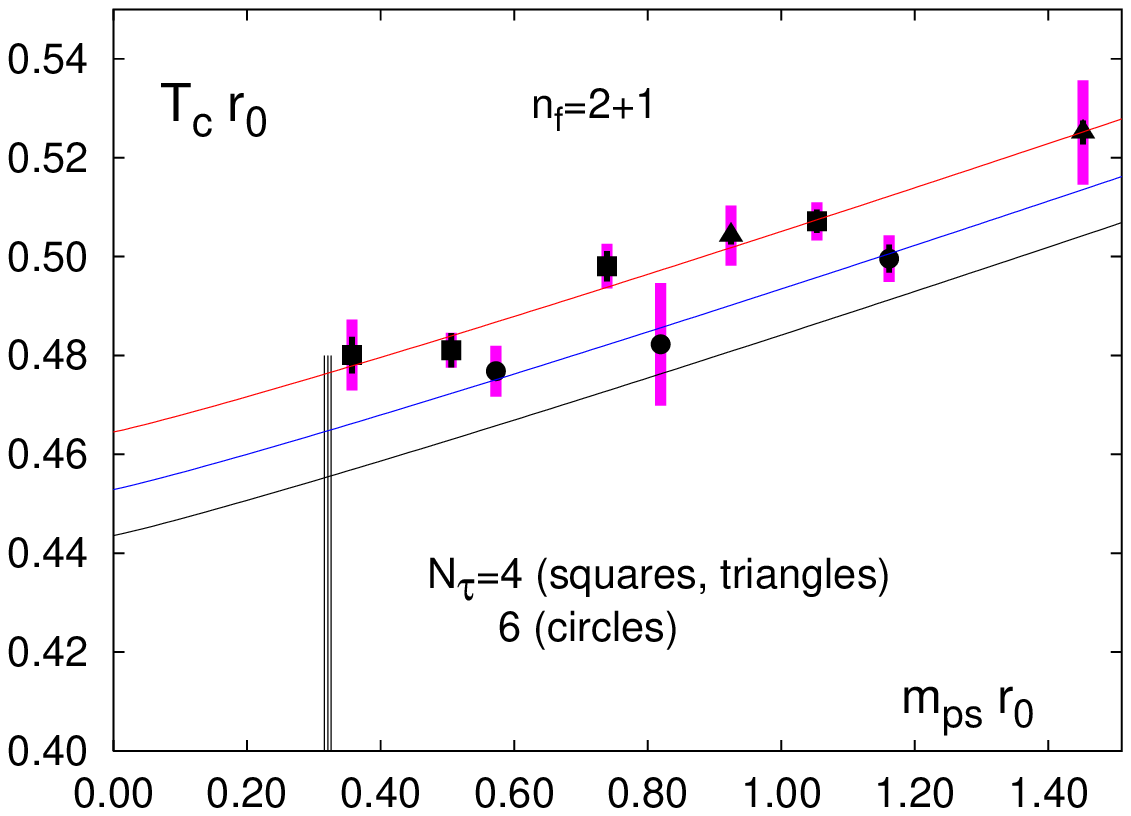, width=7.2cm}
\epsfig{file=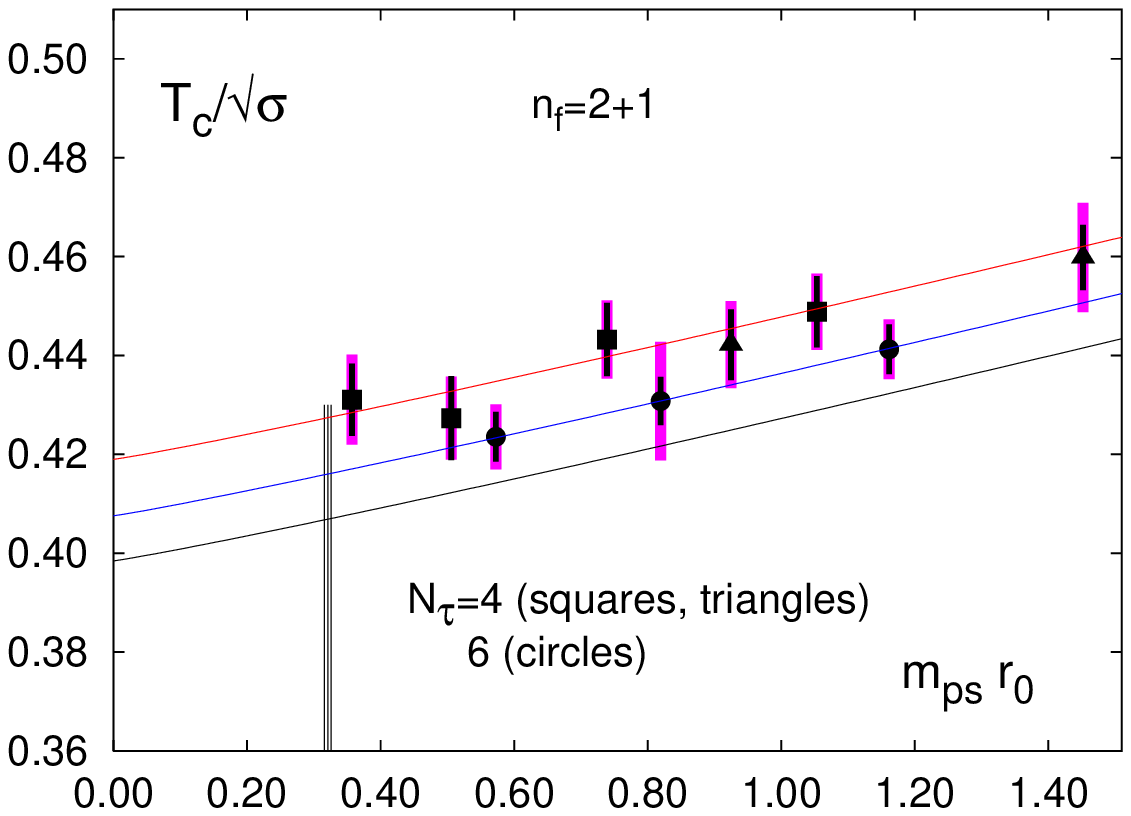, width=7.2cm}
\end{center}
\end{minipage}
\end{center}
\caption{$T_c r_0$ (left) and $T_c/\sqrt{\sigma}$ (right) as a function of 
$m_{ps}r_0$ on lattices with temporal extent $N_\tau =4$, $\hm_s= 0.065$
(squares) and $\hm_s= 0.1$ (triangles) as well as for 
$N_\tau = 6$, $\hm_s= 0.04$ (circles).
Thin error bars represent the statistical and systematic error on
$r_0/a$ and $\sqrt{\sigma}a$. The broad error bar combines this error with
the error on $\beta_c$. 
The vertical line shows the location of the physical value 
$m_{ps} r_0= 0.321(5)$ and its width represents the error on $r_0$.
The three parallel lines show results of fits based on Eq.~\ref{extrapolation}
with $d=1.08$ for $N_\tau=4,~6$ and $N_\tau \rightarrow \infty$ (top to bottom).
}
\label{fig:Tc}
\end{figure}

To obtain the transition temperature we use the results for the scales
$r_0/a$ and $\sqrt{\sigma}a$ obtained from fits to the static quark potential. 
In cases where zero temperature calculations
have not been performed directly at the critical coupling but at a nearby
$\beta$-value we use Eq.~\ref{interpolate} to determine the scales at
$\beta_c(\hm_l,\hm_s,N_\tau)$.
The transition temperature is then obtained as $T_c r_0 \equiv (r_0/a)/N_\tau$
or $T_c/\sqrt{\sigma}=1/\sqrt{\sigma} aN_\tau$.  
We show these results as function of the 
pseudo-scalar (pion) mass expressed in units of $r_0$ in Figure~\ref{fig:Tc}. 
There we give 2 errors on $T_c r_0$ and $T_c/\sqrt{\sigma}$.
A thin error bar reflects the combined statistical and systematic
errors on the scales $r_0/a$ and $\sqrt{\sigma}a$ obtained from
our 3-parameter fit to the static quark potential. 
The broad error bar combines this uncertainty of the zero temperature
scale determination with the scale-uncertainty arising from the error 
on $\beta_c$. As can be seen, the former error, which 
typically is of the order of 2\%, dominates our uncertainty on $T_cr_0$
and $T_c/\sqrt{\sigma}$ 
on the coarser $N_\tau=4$ lattices, while the uncertainty in the determination
of $\beta_c$ becomes more relevant for $N_\tau=6$. Values for the transition
temperatures are given in Table~\ref{tab:Tc}.

\begin{table}[t]
\begin{center}
\vspace{0.3cm}
\begin{tabular}{|c|l|l|c|c|}
\hline
$N_\tau$&~$\hm_s$ & $~~\hm_l$& $T_c r_0$ & $T_c/\sqrt{\sigma}$ \\
\hline
4&0.1& 0.02&   0.5043( 61)&   0.4422( 88) \\
~&~& 0.05&   0.5251(105)&   0.4598(111) \\
\hline
4&0.065& 0.00325&   0.4801( 72)&   0.4311( 91) \\
~&~& 0.0065&   0.4811( 35)&   0.4273( 84) \\
~&~& 0.013&   0.4981( 45)&   0.4432( 79) \\
~&~& 0.026&   0.5071( 38)&   0.4488( 77) \\
\hline
6&0.04& 0.004&   0.4768( 51)&   0.4236( 66) \\
~&~& 0.008&   0.4823(124)&   0.4308(120) \\
~&~& 0.016&   0.4996( 47)&   0.4413( 61) \\
\hline
\end{tabular}
\end{center}
\caption{Transition temperature in units of $r_0$ and $\sqrt{\sigma}$.
The errors given are the combined statistical errors discussed in the text.}
\label{tab:Tc}
\end{table}

The comparison of results obtained on lattices with temporal extent $N_\tau =4$
and $6$ given in Figure~\ref{fig:Tc} clearly shows   
a systematic cut-off dependence for the transition
temperature. At fixed values of $m_{ps} r_0$, results obtained for $N_\tau=6$ are 
systematically smaller than the $N_\tau =4$ results by about (3-4)\%. 
On the other
hand, we see no statistically significant dependence of our results on the value
of the strange quark mass; results obtained on the $N_\tau =4$ lattice with a
strange quark mass $\hm_s=0.1$ and $\hm_s = 0.065$ are in good agreement. The former
choice of parameters leads to a kaon mass that is about 20\% larger than in 
the latter case.

We have extrapolated our numerical results for $T_cr_0$ and $T_c/\sqrt{\sigma}$,
which have been obtained for a specific set of lattice parameters 
($\hm_l, \hm_s, N_\tau$), 
to the chiral and continuum limit using an ansatz that takes into account 
the quadratic cut-off dependence, $(aT)^2=1/N_\tau^2$, and a quark mass
dependence expressed in terms of the pseudo-scalar meson mass,
\begin{equation}
Y_{ \hm_l,\hm_s,N_\tau } = Y_{ 0,m_s,\infty } + 
A (m_{ps}r_0)^d + B/N_\tau^2 \quad,\quad Y=T_cr_0,~T_c/\sqrt{\sigma} \; ,
\label{extrapolation}
\end{equation}
If the QCD transition is second order in the chiral limit
the transition temperature
is expected to depend on the quark mass as $\hm_l^{1/\beta\delta}$, or 
correspondingly on the pseudo-scalar meson mass as $m_{ps}^{2/\beta\delta}$
with $d\equiv 2/\beta\delta\simeq 1.08$ characterizing universal scaling
behavior in the vicinity of second order phase transitions belonging to the
universality class of $O(4)$ symmetric, 3-dimensional spin models. If,
however, the transition becomes first order for small quark masses,
which is not ruled out  for physical values of the strange quark mass, 
the transition temperature
will depend linearly on the quark mass ($d=2$). A fit to our data set with
$d$ as a free fit parameter would actually favor a value smaller than unity,
although the error on $d$ is large in this case, $d= 0.6(7)$.  

Fortunately, the extrapolation to the physical point is not 
very sensitive to the choice of $d$ as our calculations have been performed
close to this point. It does, however, increase the uncertainty on
the extrapolation to the chiral limit.  
We have performed extrapolations to the chiral limit with $d$ 
varying between $d=1$ and $d=2$. From this we find

\begin{equation}
m_{ps}r_0 \equiv 0: \qquad T_c r_0 = 0.444(6)^{+12}_{-3} \quad , 
\quad T_c/\sqrt{\sigma} = 0.398(6)^{+10}_{-1} \; ,
\label{Tcchiral}
\end{equation}
where the central value is given for fits with the $O(4)$ exponent
$d=1.08$ and the lower and upper systematic error correspond to
$d=1$ and $d=2$, respectively. Using the fit values for the parameter
$A$ that controls the quark mass dependence of $T_cr_0$ ($A=0.041(5)$)
and $T_c/\sqrt{\sigma}$ ($A=0.029(4)$), respectively, we can determine
the transition temperature at the physical point, fixed by $m_{ps}r_0$r, 
where we then obtain a slightly larger value with reduced 
systematic errors,
\begin{equation}
m_{ps}r_0 \equiv 0.321(5): \qquad T_c r_0 = 0.457(7)^{+8}_{-2} \quad , 
\quad T_c/\sqrt{\sigma} = 0.408(8)^{+3}_{-1} \; .
\label{Tcphysical}
\end{equation}
Here the error includes the uncertainty in the value for the
physical point, $m_{ps}r_0$, arising from the uncertainty in the scale 
parameter $r_0 = 0.469(7)$~fm. 
We note that the extrapolated values for $T_c r_0$
and $T_c/\sqrt{\sigma}$ may also be interpreted as a continuum 
extrapolation of the shape parameters of the static potential. 
This yields  $r_0 \sqrt{\sigma} \simeq 1.11$ which is 
consistent with the continuum extrapolation 
obtained with the asqtad-action \cite{Bernard04}.

The fit parameter $B$ which controls the size of the cut-off dependent term 
in Eq.~\ref{extrapolation} is in all cases close to $1/3$. 
We find $B=0.34(9)$ for fits to $T_cr_0$ and $B=0.33(7)$ for fits to
$T_c/\sqrt{\sigma}$, respectively. The critical temperatures for $N_\tau=4$
thus are about 5\% larger than the extrapolated value, and 
for $N_\tau=6$ the difference is about 2\%. We therefore expect that 
any remaining uncertainties in our extrapolation to the continuum limit
which may arise from higher order corrections in the cut-off dependence
of $T_cr_0$ are not larger than  2\%.

The results for the transition temperature
obtained here for smaller quark masses and smaller lattice spacings is
entirely consistent with the results for 2-flavor QCD obtained previously
with the p4fat3 action on $N_\tau=4$ lattices in the chiral limit, 
$T_c/\sqrt{\sigma} =0.425(15)$ \cite{peikert}. We now find for (2+1)-flavor
QCD for $N_\tau=4$ in the chiral limit $T_c/\sqrt{\sigma} = 0.419(6)$.
The continuum extrapolated result is, however, somewhat 
larger than the continuum extrapolated result obtained
with the asqtad-action for (2+1)-flavor QCD in the chiral 
limit\footnote{In \cite{Bernard04} 
$T_c$ is given in units of $r_1$ using results for $r_1/a$ taken from
\cite{MILCpotential}. We have expressed $T_c$ in units of $r_0$ 
using $r_0/r_1=1.4795$ to convert $r_1$ to the $r_0$ scale used by us.}, 
$T_c r_0 = 0.402(29)$ \cite{Bernard04}, which is based on 
the determination of transition temperatures on lattices with temporal
extent $N_\tau =4$, $6$ and $8$.

\subsection{Using zero temperature scales to convert $T_c$ to physical units}

Although we frequently have referred to the physical value of $r_0$ during
the discussion in the previous chapters we stress
that our final result for dimensionless quantities, in particular
$T_cr_0$ and $T_c/\sqrt{\sigma}$ given in Eq.~\ref{Tcchiral}, 
does not depend on the actual physical value of $r_0$ or $\sqrt{\sigma}$.

As pointed out in the previous section, the results obtained here for $T_c$
expressed in units of $r_0$ or $\sqrt{\sigma}$ are consistent
with earlier determinations of these quantities. In fact, after
extrapolation to the continuum limit this ratio turns out to be even 
somewhat smaller than those determined previously for $2$-flavor QCD.

Unfortunately neither $r_0$ nor $\sqrt{\sigma}$ are directly measurable
experimentally. Their physical values have been deduced from lattice
calculations through a comparison with calculations for the level splitting
in the bottomonium spectrum \cite{gray,Bernard04}. 
This observable has the advantage of showing only a weak quark mass 
dependence. Of course, dealing with heavy quarks in addition to the
dynamical light quarks requires a special set-up (NRQCD) which might
introduce additional systematic errors. However, these findings 
have been cross-checked through calculations of other observables which
only involve the light quark sector. In particular, the 
pion decay constant, $f_\pi$,
has been evaluated on the MILC configurations that have
been used for the bottomonium level splitting and yields a consistent
value for $r_0$ \cite{Davies}.
One may, of course, also  consider using results for masses of mesons 
constructed from light quarks, e.g. the vector meson mass, to determine the 
scale for the transition temperature, eg. $T_c/m_\rho$. However,
even on lattices with smaller lattice spacings than those used in 
thermodynamic calculations today, the calculation of vector meson 
mass is known to suffer from large statistical and systematic 
errors \cite{MILCpotential,Davies}.
This is even more the case on the coarse lattices needed
for our finite temperature calculations. 
We thus refrained from using results on the vector meson mass for
our determination of the transition temperature.

At present the scale parameter $r_0$, deduced from the bottomonium
level splitting using NRQCD \cite{gray}, seems to be the best controlled 
lattice observable that can be used to set the scale for $T_c$.  
Using for $r_0$ the value given in Eq.~\ref{r0_fm} we obtain for the transition 
temperature in QCD at the physical point, 
\begin{equation}
T_c = 192(7)(4)\; {\rm MeV} \; ,
\label{TcMeV}
\end{equation}
where the statistical error includes the errors given in Eq.~\ref{Tcphysical}
as well as the uncertainty in the value of $r_0$ and the second error reflects
our estimate of a remaining systematic error on the extrapolation to the
continuum limit. As discussed after Eq.~\ref{Tcphysical} we estimate this error 
which arises from neglecting higher order cut-off effects in our ansatz
for the continuum extrapolation, Eq.~\ref{extrapolation} to be about 2\%.

The value of the critical temperature obtained here is about 10\% larger 
than the frequently quoted value $\sim 175$~MeV.
We note that this larger value mainly results from  
the value for $r_0$ used in our conversion to
physical scales. Together with $r_0\sqrt{\sigma} \simeq 1.11$
it implies that the string tension takes on
the value $\sqrt{\sigma} \simeq 465$~MeV. This value of the string tension
is about 10\% larger than that used in the past to set the scale for $T_c$
\cite{peikert}.

\section{Conclusions}
\label{conclusions}

We have presented new results on the transition temperature in QCD
with an almost physical quark mass spectrum. The extrapolation to the 
physical point and the continuum limit is based on numerical calculations
with an improved staggered fermion action which have been performed on
lattices with two different values of the lattice cut-off and seven different
values of bare light and strange quark masses. 

It previously has been observed that the QCD transition temperature is close
to the freeze-out temperature extracted from observed particle yields in
heavy ion experiments \cite{stachel,cleymans}. Recent results from the RHIC 
experiments determine this freeze-out temperature to be below $170$~MeV 
\cite{PHENIX,STAR}. Our results on the transition 
temperature now seem to suggest that an intermediate regime between the 
QCD transition and freeze-out exists during which the system created 
in a heavy ion collision persists in a dense hadronic phase. 

The analysis presented here leads to a value for the critical temperature with 
about 5\% statistical and systematic errors. It clearly is desirable to 
confirm our estimate of the remaining systematic errors through an additional
calculation on an even finer lattice. Furthermore, it is desirable to 
verify this result through calculations that explore other discretization
schemes for the fermion sector of QCD and to also obtain a reliable 
independent scale setting for the transition temperature from an observable 
not related to properties of the static quark potential.

\section*{Acknowledgments}
\label{ackn}
This work has been supported n part by contracts DE-AC02-98CH1-886 
and DE-FG02-92ER40699 with the U.S. Department of Energy, 
the Helmholtz Gesellschaft under grant
VI-VH-041 and the Deutsche Forschungsgemeinschaft under grant GRK 881.  
Numerical simulations have been performed
on the QCDOC computer of the RIKEN-BNL research center, the DOE funded
QCDOC at BNL and the apeNEXT at Bielefeld University.

\newpage
\begin{appendix}
\section*{Appendix: Summary of simulation parameters and results}

In this Appendix we summarize all numerical results from our simulations
with two light and a heavier strange quark mass. The header line
for all tables display the temporal lattice size $N_\tau$ and values
of the light ($\hm_l$) and strange ($\hm_s$) quark masses. The first 4 
columns of the Tables display the spatial lattice size, $N_\sigma$, the
value of the gauge coupling, $\beta$, the number of configurations and
the auto-correlation time in units of gauge field configurations generated
at the end of a Molecular Dynamics trajectory of length $\tau_{MD}=0.5$.
The next three columns give the Polyakov loop expectation value and the
light and strange quark chiral condensates. The remaining three columns
give the corresponding susceptibilities of these three observables.
Note that on the $24^3\times 6$ lattice the strange quark condensate
and its susceptibility have not been evaluated. We also do not quote
a value for the light quark chiral susceptibility in this case, as the
current statistics is not yet sufficient to determine this reliably. 
All data are given in units of appropriate powers of the lattice spacing.
%
\begin{center}
\vspace{0.3cm}
\begin{tabular}{|r|l|r|r|c|c|c|c|c|c|}
\hline
\multicolumn{9}{|c}{ $N_\tau = 4,~~\hat{m}_s = 0.065,~~\hat{m}_l=0.00325$} 
& \multicolumn{1}{c|}{ } \\
\hline
$N_\sigma$ & $~~\beta$ & \# conf. &$\tau$&
$\langle L \rangle$ &
$\langle \bar\psi \psi \rangle_l$ &
$\langle \bar\psi \psi \rangle_s$ &
$\chi_L$ & $\chi_l$ &$\chi_s$ \\
\hline
  8 &   3.2400 &   15500 &    32
 &     0.0260( 6)
 &     0.1854(17)
 &     0.2908( 9)
 &      0.086( 4)
 &       1.34( 4)
 &      0.235(13)
\\
    &   3.2600 &   19000 &    40
 &     0.0309(10)
 &     0.1647(15)
 &     0.2769( 8)
 &      0.114( 6)
 &       1.79( 7)
 &      0.289(13)
\\
    &   3.2800 &   30000 &    61
 &     0.0374( 8)
 &     0.1400(18)
 &     0.2619( 9)
 &      0.164( 7)
 &       2.53(16)
 &      0.477(41)
\\
    &   3.2900 &   30000 &    89
 &     0.0457(10)
 &     0.1189(15)
 &     0.2497( 8)
 &      0.204(10)
 &       3.01( 8)
 &      0.519(25)
\\
    &   3.3000 &   30000 &    81
 &     0.0548(13)
 &     0.0954(26)
 &     0.2372(12)
 &      0.257( 8)
 &       3.50( 9)
 &      0.631(23)
\\
    &   3.3100 &   30000 &    84
 &     0.0626(16)
 &     0.0778(31)
 &     0.2267(15)
 &      0.284(15)
 &       3.53(14)
 &      0.698(36)
\\
    &   3.3200 &   20000 &    57
 &     0.0772(10)
 &     0.0481(17)
 &     0.2098( 9)
 &      0.247(17)
 &       2.06(19)
 &      0.540(55)
\\
    &   3.3400 &   20000 &    40
 &     0.0902(12)
 &     0.0297(15)
 &     0.1939( 9)
 &      0.241(17)
 &       0.97(16)
 &      0.414(32)
\\
\hline
 16 &   3.2900 &   38960 &    66
 &     0.0424( 4)
 &     0.1313( 5)
 &     0.2513( 3)
 &      0.221(10)
 &       2.15(12)
 &      0.510(28)
\\
    &   3.3000 &   40570 &   101
 &     0.0520( 6)
 &     0.1084(10)
 &     0.2389( 5)
 &      0.279(11)
 &       3.16(17)
 &      0.645(37)
\\
    &   3.3050 &   32950 &   105
 &     0.0588( 7)
 &     0.0927(13)
 &     0.2309( 6)
 &      0.314(22)
 &       3.90(30)
 &      0.743(58)
\\
    &   3.3100 &   42300 &   102
 &     0.0649( 5)
 &     0.0791(11)
 &     0.2240( 5)
 &      0.314(11)
 &       4.12(17)
 &      0.778(34)
\\
    &   3.3200 &   39050 &    92
 &     0.0760( 4)
 &     0.0544( 8)
 &     0.2108( 4)
 &      0.310(20)
 &       3.22(21)
 &      0.663(48)
\\
\hline
\end{tabular}

\vspace{0.4cm}
Table A.1
\vspace{0.5cm}
\end{center}
\begin{center}
\vspace{0.3cm}
\begin{tabular}{|r|l|r|r|c|c|c|c|c|c|}
\hline
\multicolumn{9}{|c}{ $N_\tau = 4,~~\hat{m}_s = 0.065,~~\hat{m}_l=0.0065$} 
& \multicolumn{1}{c|}{ } \\
\hline
$N_\sigma$ & $~~\beta$ & \# conf. &$\tau$&
$\langle L \rangle$ &
$\langle \bar\psi \psi \rangle_l$ &
$\langle \bar\psi \psi \rangle_s$ &
$\chi_L$ & $\chi_l$ &$\chi_s$ \\
\hline
  8 &   3.2600 &   10000 &    37
 &     0.0272( 7)
 &     0.1868(12)
 &     0.2828( 8)
 &      0.097( 7)
 &       0.95( 9)
 &      0.289(34)
\\
    &   3.2800 &   30000 &    45
 &     0.0352( 8)
 &     0.1604(12)
 &     0.2646( 6)
 &      0.149(12)
 &       1.28(11)
 &      0.363(32)
\\
    &   3.2900 &    8900 &    56
 &     0.0394(16)
 &     0.1490(24)
 &     0.2573(12)
 &      0.146(17)
 &       1.32(19)
 &      0.334(53)
\\
    &   3.3000 &   30000 &    76
 &     0.0456( 9)
 &     0.1314(13)
 &     0.2465( 7)
 &      0.207( 5)
 &       2.11( 7)
 &      0.565(20)
\\
    &   3.3100 &   30000 &   105
 &     0.0542(21)
 &     0.1127(30)
 &     0.2346(16)
 &      0.269(18)
 &       2.51(17)
 &      0.649(54)
\\
    &   3.3200 &   34380 &   191
 &     0.0671(18)
 &     0.0869(27)
 &     0.2197(15)
 &      0.280(12)
 &       2.30( 9)
 &      0.630(32)
\\
    &   3.3300 &   30000 &   101
 &     0.0780(17)
 &     0.0665(25)
 &     0.2067(14)
 &      0.288(26)
 &       1.99(24)
 &      0.626(70)
\\
    &   3.3400 &   20000 &    75
 &     0.0884(18)
 &     0.0506(23)
 &     0.1957(16)
 &      0.248(14)
 &       1.13(11)
 &      0.472(35)
\\
    &   3.3600 &   12750 &    30
 &     0.1017(25)
 &     0.0344(14)
 &     0.1798(16)
 &      0.242(16)
 &       0.50( 7)
 &      0.361(29)
\\
\hline
 16 &   3.2800 &   20510 &    56
 &     0.0312( 3)
 &     0.1662( 3)
 &     0.2664( 2)
 &      0.168(10)
 &       1.10( 9)
 &      0.379(23)
\\
    &   3.2900 &   30160 &    79
 &     0.0380( 6)
 &     0.1507( 8)
 &     0.2562( 4)
 &      0.212(16)
 &       1.38(10)
 &      0.432(29)
\\
    &   3.3000 &   36100 &    76
 &     0.0445( 3)
 &     0.1351( 6)
 &     0.2464( 3)
 &      0.244(16)
 &       1.96(15)
 &      0.574(42)
\\
    &   3.3100 &   40440 &   110
 &     0.0542( 4)
 &     0.1146( 5)
 &     0.2343( 2)
 &      0.316(14)
 &       2.74(11)
 &      0.740(34)
\\
    &   3.3150 &   45570 &   141
 &     0.0612( 6)
 &     0.1007(10)
 &     0.2262( 6)
 &      0.334(15)
 &       3.12(17)
 &      0.820(44)
\\
    &   3.3200 &   32310 &    81
 &     0.0666( 7)
 &     0.0896(10)
 &     0.2198( 5)
 &      0.304(17)
 &       2.51(20)
 &      0.647(51)
\\
\hline
\end{tabular}

\vspace{0.4cm}
Table A.2
\vspace{0.5cm}
\end{center}
\begin{center}
\vspace{0.3cm}
\begin{tabular}{|r|l|r|r|c|c|c|c|c|c|}
\hline
\multicolumn{9}{|c}{ $N_\tau = 4,~~\hat{m}_s = 0.065,~~\hat{m}_l=0.013$} 
& \multicolumn{1}{c|}{ } \\
\hline
$N_\sigma$ & $~~\beta$ & \# conf. &$\tau$&
$\langle L \rangle$ &
$\langle \bar\psi \psi \rangle_l$ &
$\langle \bar\psi \psi \rangle_s$ &
$\chi_L$ & $\chi_l$ &$\chi_s$ \\
\hline
  8 &   3.2600 &   10000 &    26
 &     0.0255( 9)
 &     0.2075( 7)
 &     0.2865( 6)
 &      0.083( 7)
 &       0.48( 3)
 &      0.228( 9)
\\
    &   3.2800 &   10000 &    45
 &     0.0315( 9)
 &     0.1865(19)
 &     0.2705(14)
 &      0.118( 8)
 &       0.73( 7)
 &      0.314(31)
\\
    &   3.3000 &   20000 &    58
 &     0.0383(13)
 &     0.1645(20)
 &     0.2545(13)
 &      0.168(11)
 &       1.13(11)
 &      0.436(44)
\\
    &   3.3200 &   30000 &   109
 &     0.0520(18)
 &     0.1372(22)
 &     0.2355(14)
 &      0.254(14)
 &       1.64(10)
 &      0.621(38)
\\
    &   3.3300 &   30000 &   134
 &     0.0621(18)
 &     0.1188(22)
 &     0.2231(14)
 &      0.294(12)
 &       1.68( 9)
 &      0.620(38)
\\
    &   3.3400 &   20000 &    79
 &     0.0747(18)
 &     0.0991(25)
 &     0.2098(16)
 &      0.303(15)
 &       1.68( 9)
 &      0.669(38)
\\
    &   3.3600 &   17720 &    52
 &     0.0948(17)
 &     0.0695(19)
 &     0.1879(13)
 &      0.268(14)
 &       0.96( 7)
 &      0.473(21)
\\
    &   3.3800 &   10000 &    40
 &     0.1091(15)
 &     0.0521(10)
 &     0.1707( 8)
 &      0.237(11)
 &       0.41( 3)
 &      0.296(20)
\\
\hline
 16 &   3.3200 &   20680 &    85
 &     0.0501( 6)
 &     0.1387( 8)
 &     0.2356( 5)
 &      0.293(24)
 &       1.60(16)
 &      0.612(66)
\\
    &   3.3250 &   54840 &   114
 &     0.0554( 6)
 &     0.1295( 7)
 &     0.2295( 4)
 &      0.298(16)
 &       1.82(14)
 &      0.680(51)
\\
    &   3.3300 &   50000 &   149
 &     0.0615( 5)
 &     0.1185( 7)
 &     0.2222( 4)
 &      0.340(13)
 &       1.94( 8)
 &      0.730(33)
\\
    &   3.3350 &   55600 &   124
 &     0.0673( 4)
 &     0.1091( 6)
 &     0.2160( 4)
 &      0.330(17)
 &       1.92(10)
 &      0.711(37)
\\
    &   3.3400 &   60000 &   104
 &     0.0741( 6)
 &     0.0980( 8)
 &     0.2085( 5)
 &      0.355(12)
 &       1.95( 7)
 &      0.733(26)
\\
    &   3.3450 &   32560 &    82
 &     0.0803( 5)
 &     0.0886( 6)
 &     0.2021( 4)
 &      0.315(18)
 &       1.61(13)
 &      0.650(47)
\\
    &   3.3500 &   20780 &    65
 &     0.0859( 5)
 &     0.0801( 6)
 &     0.1960( 4)
 &      0.303(20)
 &       1.31(11)
 &      0.572(46)
\\
\hline
\end{tabular}

\vspace{0.4cm}
Table A.3
\vspace{0.5cm}
\end{center}
\begin{center}
\vspace{0.3cm}
\begin{tabular}{|r|l|r|r|c|c|c|c|c|c|}
\hline
\multicolumn{9}{|c}{ $N_\tau = 4,~~\hat{m}_s = 0.065,~~\hat{m}_l=0.026$} 
& \multicolumn{1}{c|}{ } \\
\hline
$N_\sigma$ & $~~\beta$ & \# conf. &$\tau$&
$\langle L \rangle$ &
$\langle \bar\psi \psi \rangle_l$ &
$\langle \bar\psi \psi \rangle_s$ &
$\chi_L$ & $\chi_l$ &$\chi_s$ \\
\hline
  8 &   3.2600 &   10000 &    30
 &     0.0235( 6)
 &     0.2378(14)
 &     0.2918(10)
 &      0.076( 5)
 &       0.35( 4)
 &      0.217(21)
\\
    &   3.2800 &   10000 &    22
 &     0.0270( 9)
 &     0.2241( 8)
 &     0.2802( 5)
 &      0.096( 7)
 &       0.38( 4)
 &      0.226(20)
\\
    &   3.3000 &    9530 &    39
 &     0.0311(12)
 &     0.2091(15)
 &     0.2675(12)
 &      0.134(10)
 &       0.58( 5)
 &      0.339(28)
\\
    &   3.3200 &    9270 &    66
 &     0.0387(22)
 &     0.1884(25)
 &     0.2507(19)
 &      0.161(14)
 &       0.66(10)
 &      0.381(56)
\\
    &   3.3400 &   20000 &    64
 &     0.0484(16)
 &     0.1673(16)
 &     0.2335(11)
 &      0.237(16)
 &       0.91( 6)
 &      0.496(37)
\\
    &   3.3500 &   30000 &   118
 &     0.0580(23)
 &     0.1520(26)
 &     0.2217(19)
 &      0.285(15)
 &       1.09(10)
 &      0.594(54)
\\
    &   3.3600 &   30000 &   128
 &     0.0737(21)
 &     0.1315(24)
 &     0.2057(18)
 &      0.344(21)
 &       1.26( 8)
 &      0.697(43)
\\
    &   3.3700 &   30000 &    91
 &     0.0819(23)
 &     0.1189(22)
 &     0.1955(16)
 &      0.342(33)
 &       1.16(15)
 &      0.673(77)
\\
    &   3.3800 &   20000 &   107
 &     0.0965(21)
 &     0.1033(20)
 &     0.1825(16)
 &      0.338(23)
 &       0.95( 8)
 &      0.594(46)
\\
    &   3.4000 &   10000 &    39
 &     0.1108(16)
 &     0.0862(12)
 &     0.1672( 9)
 &      0.253(24)
 &       0.48( 7)
 &      0.346(44)
\\
\hline
 16 &   3.3000 &    9050 &    32
 &     0.0255( 3)
 &     0.2105( 3)
 &     0.2684( 2)
 &      0.138( 8)
 &       0.46( 3)
 &      0.283(25)
\\
    &   3.3100 &    6890 &    42
 &     0.0291( 6)
 &     0.2012( 6)
 &     0.2607( 4)
 &      0.162(12)
 &       0.49( 5)
 &      0.296(36)
\\
    &   3.3350 &   16870 &   117
 &     0.0422( 8)
 &     0.1756( 9)
 &     0.2399( 7)
 &      0.235(20)
 &       0.82( 8)
 &      0.453(41)
\\
    &   3.3500 &   26500 &    85
 &     0.0569( 5)
 &     0.1529( 5)
 &     0.2221( 3)
 &      0.307(18)
 &       1.14( 7)
 &      0.608(39)
\\
    &   3.3550 &   38760 &   110
 &     0.0626( 6)
 &     0.1446( 6)
 &     0.2156( 4)
 &      0.334(15)
 &       1.15( 5)
 &      0.618(30)
\\
    &   3.3600 &   29780 &   215
 &     0.0681(10)
 &     0.1370(11)
 &     0.2097( 8)
 &      0.443(20)
 &       1.69(11)
 &      0.910(65)
\\
    &   3.3625 &   37880 &   101
 &     0.0729( 5)
 &     0.1312( 4)
 &     0.2052( 3)
 &      0.371(24)
 &       1.25(10)
 &      0.678(53)
\\
    &   3.3650 &   40000 &   101
 &     0.0757( 7)
 &     0.1272( 7)
 &     0.2020( 5)
 &      0.359(27)
 &       1.26(12)
 &      0.691(64)
\\
    &   3.3700 &   60000 &    85
 &     0.0833( 3)
 &     0.1179( 3)
 &     0.1947( 3)
 &      0.359(13)
 &       1.17( 5)
 &      0.664(31)
\\
    &   3.3750 &   60000 &    75
 &     0.0903( 3)
 &     0.1095( 3)
 &     0.1879( 3)
 &      0.332(15)
 &       0.98( 4)
 &      0.577(27)
\\
\hline
\end{tabular}

\vspace{0.4cm}
Table A.4
\vspace{0.5cm}
\end{center}
\begin{center}
\vspace{0.3cm}
\begin{tabular}{|r|l|r|r|c|c|c|c|c|c|}
\hline
\multicolumn{9}{|c}{ $N_\tau = 4,~~\hat{m}_s = 0.1,~~\hat{m}_l=0.05$} 
& \multicolumn{1}{c|}{ } \\
\hline
$N_\sigma$ & $~~\beta$ & \# conf. &$\tau$&
$\langle L \rangle$ &
$\langle \bar\psi \psi \rangle_l$ &
$\langle \bar\psi \psi \rangle_s$ &
$\chi_L$ & $\chi_l$ &$\chi_s$ \\
\hline
  8 &   3.3600 &    6900 &    29
 &     0.0326(18)
 &     0.2295(13)
 &     0.2939(10)
 &      0.137(16)
 &       0.29( 3)
 &      0.172(20)
\\
    &   3.3800 &    6900 &    69
 &     0.0424(36)
 &     0.2106(27)
 &     0.2784(20)
 &      0.246(34)
 &       0.55(11)
 &      0.317(65)
\\
    &   3.4000 &   27740 &   105
 &     0.0542(16)
 &     0.1919(15)
 &     0.2633(11)
 &      0.321(16)
 &       0.71( 3)
 &      0.402(17)
\\
    &   3.4200 &   59900 &   114
 &     0.0786(17)
 &     0.1656(13)
 &     0.2423(10)
 &      0.369(15)
 &       0.80( 4)
 &      0.463(25)
\\
    &   3.4350 &   59290 &   148
 &     0.0934(27)
 &     0.1498(19)
 &     0.2293(15)
 &      0.405(20)
 &       0.69( 4)
 &      0.411(24)
\\
    &   3.4500 &   38450 &    77
 &     0.1120(19)
 &     0.1336(11)
 &     0.2158( 9)
 &      0.324(26)
 &       0.46( 4)
 &      0.298(23)
\\
    &   3.4750 &    7000 &   120
 &     0.1280(48)
 &     0.1195(30)
 &     0.2027(25)
 &      0.399(74)
 &       0.40(14)
 &      0.291(89)
\\
    &   3.5000 &    1400 &    17
 &     0.1511(11)
 &     0.1045(14)
 &     0.1873(13)
 &      0.151(44)
 &       0.08( 2)
 &      0.073(20)
\\
\hline
\end{tabular}

\vspace{0.4cm}
Table A.5
\vspace{0.5cm}
\end{center}
\begin{center}
\vspace{0.3cm}
\begin{tabular}{|r|l|r|r|c|c|c|c|c|c|}
\hline
\multicolumn{9}{|c}{ $N_\tau = 4,~~\hat{m}_s = 0.1,~~\hat{m}_l=0.02$} 
& \multicolumn{1}{c|}{ } \\
\hline
$N_\sigma$ & $~~\beta$ & \# conf. &$\tau$&
$\langle L \rangle$ &
$\langle \bar\psi \psi \rangle_l$ &
$\langle \bar\psi \psi \rangle_s$ &
$\chi_L$ & $\chi_l$ &$\chi_s$ \\
\hline
  8 &   3.3200 &    6250 &    33
 &     0.0320(17)
 &     0.1873(21)
 &     0.3016(11)
 &      0.104(11)
 &       0.45( 5)
 &      0.166(14)
\\
    &   3.3400 &   29120 &   102
 &     0.0444(19)
 &     0.1622(20)
 &     0.2852(11)
 &      0.202(16)
 &       0.92(11)
 &      0.303(32)
\\
    &   3.3600 &   49210 &    99
 &     0.0613(11)
 &     0.1334(10)
 &     0.2667( 5)
 &      0.314(17)
 &       1.42( 7)
 &      0.452(24)
\\
    &   3.3800 &   30000 &   183
 &     0.0835(18)
 &     0.1016(20)
 &     0.2461(11)
 &      0.344(13)
 &       1.24( 8)
 &      0.436(27)
\\
    &   3.4000 &    6300 &    37
 &     0.1031(25)
 &     0.0770(20)
 &     0.2284(14)
 &      0.259(23)
 &       0.50( 5)
 &      0.233(16)
\\
    &   3.4200 &    6500 &    21
 &     0.1149(19)
 &     0.0650(11)
 &     0.2171( 8)
 &      0.240(17)
 &       0.26( 3)
 &      0.173(17)
\\
\hline
\end{tabular}

\vspace{0.4cm}
Table A.6
\vspace{0.5cm}
\end{center}
\begin{center}
\vspace{0.3cm}
\begin{tabular}{|r|l|r|r|c|c|c|c|c|c|}
\hline
\multicolumn{9}{|c}{ $N_\tau = 6,~~\hat{m}_s = 0.04,~~\hat{m}_l=0.004$} 
& \multicolumn{1}{c|}{ } \\
\hline
$N_\sigma$ & $~~\beta$ & \# conf. &$\tau$&
$\langle L \rangle$ &
$\langle \bar\psi \psi \rangle_l$ &
$\langle \bar\psi \psi \rangle_s$ &
$\chi_L$ & $\chi_l$ &$\chi_s$ \\
\hline
 16 &   3.4400 &   25850 &    75
 &     0.0146( 2)
 &     0.0553( 5)
 &     0.1371( 3) 
 &      0.106( 6)
 &       1.10( 4)
 &      0.337(14)
\\
    &   3.4500 &   38680 &    77
 &     0.0182( 3)
 &     0.0451( 6)
 &     0.1301( 3) 
 &      0.128( 6)
 &       1.36(11)
 &      0.380(31)
\\
    &   3.4550 &   40030 &    65
 &     0.0191( 3)
 &     0.0417( 5)
 &     0.1274( 3) 
 &      0.133( 4)
 &       1.30( 6)
 &      0.362(22)
\\
    &   3.4600 &   60000 &    97
 &     0.0216( 3)
 &     0.0361( 3)
 &     0.1236( 2) 
 &      0.153( 7)
 &       1.57(11)
 &      0.455(34)
\\
    &   3.4650 &   40030 &   129
 &     0.0232( 4)
 &     0.0326( 7)
 &     0.1208( 4) 
 &      0.162( 8)
 &       1.48( 8)
 &      0.438(19)
\\
    &   3.4700 &   30000 &    90
 &     0.0254( 5)
 &     0.0287( 7)
 &     0.1179( 4) 
 &      0.155( 9)
 &       1.16( 9)
 &      0.410(35)
\\
    &   3.4800 &   30000 &    50
 &     0.0298( 4)
 &     0.0218( 3)
 &     0.1119( 2) 
 &      0.149( 6)
 &       0.62( 5)
 &      0.300(29)
\\
\hline
 24 &   3.4450 &    5750 &    71
 &     0.0143( 4)
 &     0.0530( 2)
 &   -
 &      0.111(10)
 &   -
 &   -
\\
    &   3.4500 &    8110 &    52
 &     0.0178( 4)
 &     0.0453( 2)
 &   -
 &      0.150( 8)
 &   -
 &   -
\\
    &   3.4550 &    6780 &    34
 &     0.0199( 3)
 &     0.0402( 2)
 &   -
 &      0.115( 6)
 &   -
 &   -
\\
    &   3.4600 &    5240 &    40
 &     0.0206( 4)
 &     0.0369( 2)
 &   -
 &      0.131(10)
 &   -
 &   -
\\
    &   3.4650 &    6830 &    73
 &     0.0239( 5)
 &     0.0313( 3)
 &   -
 &      0.159(21)
 &   -
 &   -
\\
    &   3.4700 &    5760 &    86
 &     0.0258( 6)
 &     0.0277( 2)
 &   -
 &      0.155(14)
 &   -
 &   -
\\
\hline
\end{tabular}

\vspace{0.4cm}
Table A.7
\vspace{0.5cm}
\end{center}
\begin{center}
\vspace{0.3cm}
\begin{tabular}{|r|l|r|r|c|c|c|c|c|c|}
\hline
\multicolumn{9}{|c}{ $N_\tau = 6,~~\hat{m}_s = 0.04,~~\hat{m}_l=0.008$} 
& \multicolumn{1}{c|}{ } \\
\hline
$N_\sigma$ & $~~\beta$ & \# conf. &$\tau$&
$\langle L \rangle$ &
$\langle \bar\psi \psi \rangle_l$ &
$\langle \bar\psi \psi \rangle_s$ &
$\chi_L$ & $\chi_l$ &$\chi_s$ \\
\hline
 16 &   3.4500 &   51200 &    85
 &     0.0144( 2)
 &     0.0661( 2)
 &     0.1349( 1)
 &      0.116( 5)
 &       0.92( 8)
 &      0.392(32)
\\
    &   3.4600 &   30980 &    80
 &     0.0174( 4)
 &     0.0582( 5)
 &     0.1286( 3)
 &      0.135( 7)
 &       1.05( 7)
 &      0.430(30)
\\
    &   3.4650 &   53730 &   128
 &     0.0194( 3)
 &     0.0536( 4)
 &     0.1251( 3)
 &      0.149( 7)
 &       1.10( 9)
 &      0.445(39)
\\
    &   3.4700 &   62490 &    64
 &     0.0215( 2)
 &     0.0495( 3)
 &     0.1219( 2)
 &      0.156( 6)
 &       1.00( 5)
 &      0.398(21)
\\
    &   3.4750 &   59950 &    94
 &     0.0237( 4)
 &     0.0452( 5)
 &     0.1185( 3)
 &      0.166( 7)
 &       1.08( 7)
 &      0.440(27)
\\
    &   3.4800 &   26670 &    52
 &     0.0253( 4)
 &     0.0422( 5)
 &     0.1159( 3)
 &      0.168( 9)
 &       0.92( 8)
 &      0.384(34)
\\
    &   3.4900 &   18080 &    42
 &     0.0297( 4)
 &     0.0355( 5)
 &     0.1102( 3)
 &      0.155( 7)
 &       0.56( 5)
 &      0.289(22)
\\
    &   3.5000 &   13190 &    29
 &     0.0323( 5)
 &     0.0314( 3)
 &     0.1060( 2)
 &      0.155( 7)
 &       0.37( 3)
 &      0.223(11)
\\
    &   3.5100 &   10350 &    23
 &     0.0361( 6)
 &     0.0280( 4)
 &     0.1020( 4)
 &      0.133(10)
 &       0.30( 3)
 &      0.222(25)
\\
\hline
 32 &   3.4700 &   18240 &    92
 &     0.0211( 2)
 &     0.0496( 3)
 &     0.1219( 2)
 &      0.149(10)
 &       0.99(10)
 &      0.393(41)
\\
\hline
\end{tabular}

\vspace{0.4cm}
Table A.8
\vspace{0.5cm}
\end{center}
\begin{center}
\vspace{0.3cm}
\begin{tabular}{|r|l|r|r|c|c|c|c|c|c|}
\hline
\multicolumn{9}{|c}{ $N_\tau = 6,~~\hat{m}_s = 0.04,~~\hat{m}_l=0.016$} 
& \multicolumn{1}{c|}{ } \\
\hline
$N_\sigma$ & $~~\beta$ & \# conf. &$\tau$&
$\langle L \rangle$ &
$\langle \bar\psi \psi \rangle_l$ &
$\langle \bar\psi \psi \rangle_s$ &
$\chi_L$ & $\chi_l$ &$\chi_s$ \\
\hline
 16 &   3.4200 &   10000 &    34
 &     0.0079( 2)
 &     0.1137( 3)
 &     0.1589( 3)
 &      0.054( 2)
 &       0.42( 3)
 &      0.288(19)
\\
    &   3.4300 &   10000 &    40
 &     0.0083( 1)
 &     0.1076( 3)
 &     0.1533( 3)
 &      0.060( 3)
 &       0.37( 4)
 &      0.251(31)
\\
    &   3.4400 &   10000 &    38
 &     0.0091( 3)
 &     0.1009( 4)
 &     0.1472( 4)
 &      0.068( 4)
 &       0.36( 4)
 &      0.237(22)
\\
    &   3.4500 &   10000 &    28
 &     0.0104( 3)
 &     0.0948( 4)
 &     0.1417( 3)
 &      0.090( 6)
 &       0.40( 3)
 &      0.256(18)
\\
    &   3.4600 &   10000 &    43
 &     0.0124( 5)
 &     0.0879( 7)
 &     0.1355( 5)
 &      0.104( 6)
 &       0.52( 5)
 &      0.329(29)
\\
    &   3.4700 &   18410 &    49
 &     0.0152( 4)
 &     0.0812( 5)
 &     0.1296( 4)
 &      0.112( 5)
 &       0.55( 3)
 &      0.336(20)
\\
    &   3.4800 &   11390 &    41
 &     0.0178( 6)
 &     0.0743( 5)
 &     0.1235( 4)
 &      0.132(10)
 &       0.49( 4)
 &      0.301(24)
\\
    &   3.4900 &   18920 &    49
 &     0.0220( 4)
 &     0.0681( 4)
 &     0.1181( 3)
 &      0.166(12)
 &       0.62( 6)
 &      0.357(30)
\\
    &   3.5000 &   20000 &    81
 &     0.0269( 7)
 &     0.0605( 6)
 &     0.1115( 5)
 &      0.200(10)
 &       0.72( 7)
 &      0.426(37)
\\
    &   3.5100 &   13510 &    62
 &     0.0312( 9)
 &     0.0551( 8)
 &     0.1065( 6)
 &      0.173(15)
 &       0.48( 7)
 &      0.308(42)
\\
    &   3.5200 &    8640 &    27
 &     0.0350( 5)
 &     0.0503( 4)
 &     0.1020( 3)
 &      0.154(12)
 &       0.31( 3)
 &      0.212(19)
\\
\hline
\end{tabular}

\vspace{0.4cm}
Table A.9
\vspace{0.5cm}
\end{center}
\end{appendix}

\end{document}